\documentclass[twocolumn, 10pt, aps, floatfix, showpacs, prl,
  citeautoscript, superscriptaddress]{revtex4-1} 
\usepackage{graphicx}
\usepackage[caption=false]{subfig}
\usepackage{amsmath, amssymb}
\usepackage{bm}
\usepackage{xcolor}
\usepackage{upgreek}
\usepackage[colorlinks, citecolor={blue!50!black}, urlcolor={blue!50!black}, linkcolor={red!50!black}]{hyperref}
\usepackage{bookmark}
\usepackage[version=4]{mhchem}
\usepackage{microtype}
\usepackage{colordvi}
\usepackage[load=physical,load=abbr, range-units=single]{siunitx}

\newcommand{\pmat}[1]{\begin{pmatrix}#1\end{pmatrix}}
\newcommand{\avg}[1]{\langle#1\rangle}
\newcommand{\comment}[1]{}

\def\e{\varepsilon}
\renewcommand{\comment}{\paragraph}
\begin{document}
\title{Platform for nodal topological superconductors in monolayer molybdenum dichalcogenides}

\author{Lin Wang}
\email[Corresponding author: ]{L.Wang-6@tudelft.nl}
\affiliation{Kavli Institute of Nanoscience, Delft University of Technology,
  P.~O.~Box 4056, 2600 GA Delft, The Netherlands}
\author{Tomas Orn Rosdahl}
\affiliation{Kavli Institute of Nanoscience, Delft University of Technology,
  P.~O.~Box 4056, 2600 GA Delft, The Netherlands}
\author{Doru Sticlet}
\affiliation{National  Institute  for  Research  and  Development of Isotopic and Molecular  Technologies,  67-103 Donat, 400293 Cluj-Napoca, Romania}
\affiliation{Kavli Institute of Nanoscience, Delft University of Technology, P.~O.~Box 4056, 2600 GA Delft, The Netherlands}

\begin{abstract}
We propose a platform to realize nodal topological superconductors in a superconducting monolayer of \ce{MoX2} (X$=$S, Se, Te) using an in-plane magnetic field.
The bulk nodal points appear where the spin splitting due to spin-orbit coupling
vanishes near the $\pm \bm K$ valleys of the Brillouin zone, and are six or twelve per valley in total.
In the nodal topological superconducting phase, the nodal points are connected
by flat bands of zero-energy Andreev edge states. These flat bands, which are
protected by chiral symmetry, are present for all lattice-termination boundaries except zigzag.
\end{abstract}

\maketitle

\section{Introduction}
Fully gapped topological superconductors (TSCs), characterized by a global topological invariant in the Brillouin zone, have been the subject of intense investigation in recent years.
They provide a platform for the creation of the Majorana
quasiparticle~\cite{Alicea2012, Leijnse2012, Beenakker2013}, which has promising
applications in quantum information~\cite{Kitaev2001, Bravyi2002, Nayak2008}.
Nodal superconductors, {i.e.},
superconductors with nodal points or lines at the Fermi surface where the bulk gap vanishes, can also display nontrivial topological properties, becoming nodal
TSCs~\cite{Kashiwaya2000, Lofwander2001, Schnyder2015}.
Their topological invariants are only defined locally in the Brillouin zone, giving rise to flat bands or arcs of surface states in the nontrivial phase \cite{Ryu2002, Sato2011, Schnyder2012}.

Intrinsic nodal TSCs are predicted to exist in unconventional superconductors,
such as high-temperature $d$-wave superconductors~\cite{Tsuei2000}, the heavy
fermion systems~\cite{Kasahara2007, Zhou2013,
  Allan2013}, noncentrosymmetric superconductors~\cite{Yada2011, Schnyder2011},
and Weyl superconductors~\cite{Fischer2014}. However, intrinsic unconventional
pairing is complex and ambiguous, and is furthermore not robust to disorder, making intrinsic nodal TSCs challenging experimentally.
It is therefore desirable to engineer nodal TSCs using simpler components \cite{Meng2012, Wong2013, Huang2018}, such as conventional $s$-wave spin-singlet superconductors, similar
to efforts in realizing fully gapped TSCs using proximity-induced $s$-wave pairing~\cite{Fu2008, Sau2010}.

Two-dimensional monolayers of transition metal dichalcogenides (TMDs)~\cite{Wang2012} offer an opportunity to engineer nodal TSCs.
Recent experiments show that several monolayer TMDs, such as \ce{MoS2}, \ce{MoSe2}, \ce{MoTe2}, \ce{WS2}, and \ce{NbSe2}, become superconducting~\cite{Ye2012, Taniguchi2012, Lu2015, Shi2015, Saito2016, Costanzo2016, Xi2016, Zheliuk2017, Sergio2018}, with a critical temperature, {e.g.,}~as large as $10$ K observed in \ce{MoS2}~\cite{Lu2015}.
These superconductors possess an extremely high critical in-plane magnetic
field, several times larger than the Pauli limit, due to a special type of Ising
spin-orbit coupling (SOC)~\cite{Lu2015, Saito2016, Zhou2016, Ilic2017}.
The Ising SOC results from the heavy atoms and the absence of inversion symmetry, and acts as an effective Zeeman term perpendicular to the TMD plane, with opposite orientation at opposite momenta, pinning electron spins to the out-of-plane direction~\cite{Zhu2011, Xiao2012}.
Previous work predicts that hole-doped monolayer \ce{NbSe2} with $s$-wave
superconductivity near $\bm\Gamma$ becomes a nodal TSC in an in-plane magnetic
field~\cite{He2016}. In their proposal, the bulk nodal points appear along
  $\Gamma-M$ lines where the Ising SOC vanishes because of the in-plane mirror
symmetry $M_x: x\rightarrow -x$.
However, the potential of TMD materials such as \ce{MoS2}, \ce{MoSe2},
\ce{MoTe2}, and \ce{WS2}, which are superconducting at electron doping near the
$\bm K$ valleys, to become nodal TSC, is currently not known. Note that $M_x$ does not guarantee the vanishing of SOC near the $\bm K$ valleys.

In this paper, we show that $s$-wave superconducting monolayers of molybdenum dichalcogenides (\ce{MoX2}, X$=$S, Se, Te) become nodal TSCs in the presence of an in-plane magnetic field.
In this previously unknown topological phase, the bulk nodal points appear near the $\bm K$ valleys at special momenta where the SOC splitting vanishes.
We find two regimes in the nodal topological phase, with six or twelve nodal points appearing near each valley respectively.
In the nodal topological phase, nodal points are connected by zero-energy Andreev flat band edge states, which are protected by a chiral symmetry originating from mirror symmetry in the \ce{MoX2} plane, and present for all edges except zigzag.
Finally, we discuss possible experimental verification of the nodal topological phase.
\nocite{Roldan2014, Kane2005, Slater1954, Peierls1933, Hofstadter1976, Boykin2001}

\section{Model}
\begin{figure}[!tbh]
\includegraphics[width=0.97\columnwidth]{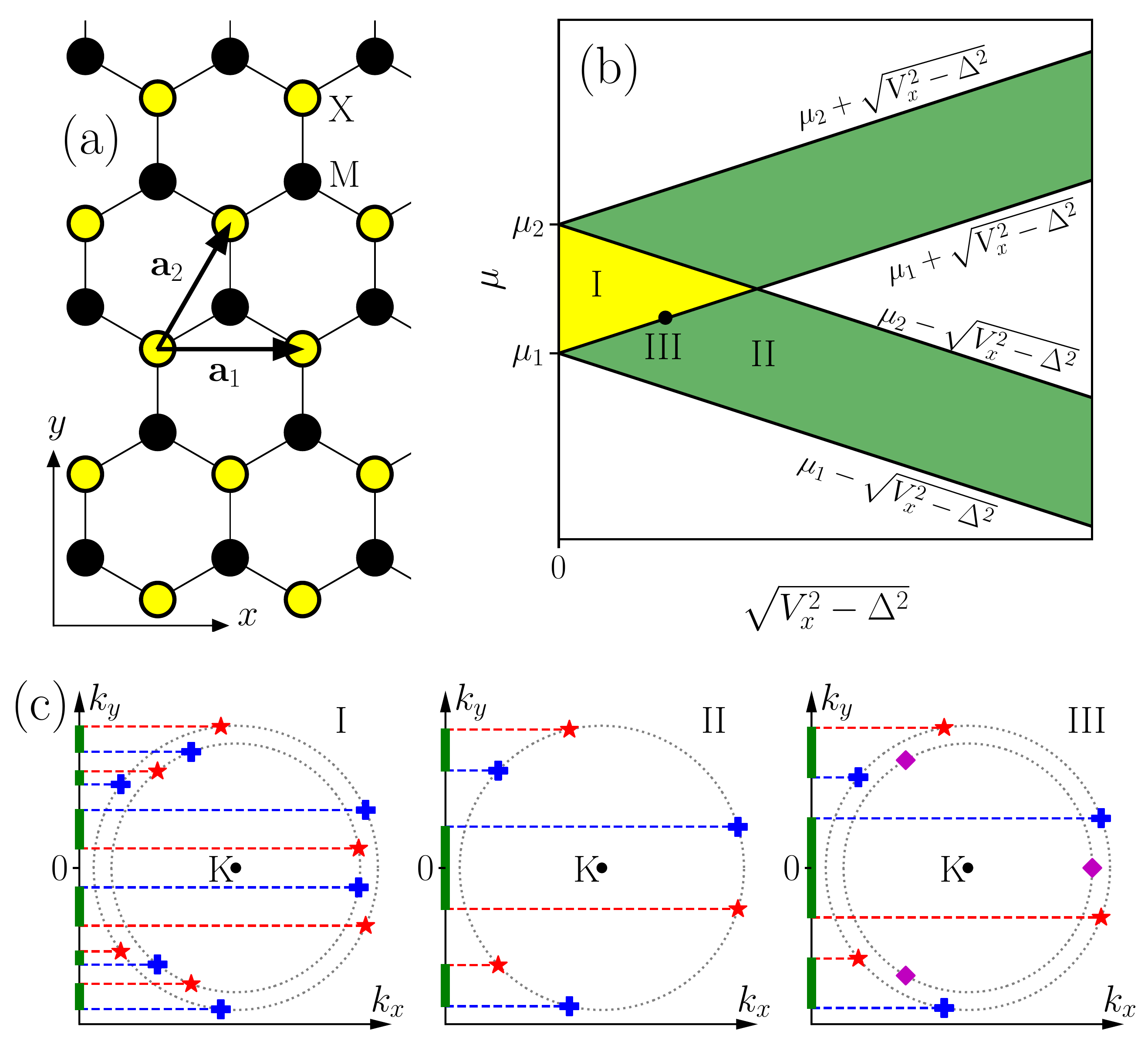}
\caption{(a) Top view of monolayer \ce{MX2} lattice structure with primitive lattice vectors ${\bm a}_1$ and
  ${\bm a}_2$.
  (b) Phase diagram of the gap-closing condition as a
  function of $\mu$ and $\sqrt{V_x^2-\Delta^2}$.
  Nodal points appear in regions where the gap closes, colored yellow (regime I) and green
  (regime II),
  with the phase boundaries given by $\mu=\mu_{1,2}\pm
  \sqrt{V_x^2-\Delta^2}$. III represents the boundary between regimes I and II. (c) Sketch of nodal points near $\bm K$ valley. 
The chirality of nodal points with $+ (\star)$ is $1 (-1)$, and diamond denotes two overlapping nodal points of opposite chirality.
Nodal point projections on the $k_y$-axis determine topologically nontrivial phases with nonzero winding number (solid green lines).}
\label{fig:hex_nodal_points}
\end{figure}

A monolayer \ce{MX2} (\ce{MoS2}, \ce{MoSe2}, \ce{MoTe2}, or \ce{WS2}) consists of a triangular lattice of M atoms sandwiched between two layers of X atoms, each also forming a triangular lattice.
The top and bottom X atoms project onto the same position in the layer of M atoms, such that when viewed from above, the monolayer has the hexagonal lattice structure shown in Fig.~\ref{fig:hex_nodal_points}(a), with primitive lattice vectors ${\bm a}_1$ and ${\bm a}_2$.
In the normal state, the monolayer \ce{MX2} has a direct band gap at the $\pm \bm K$ points.
Near the $\eta \bm K$ ($\eta = \pm$) points, the point group is $C_{3h}$, and the effective Hamiltonian of the lowest
conduction band up to the third order in momentum ${\bm k} = (k_x, k_y)$ is
\begin{equation}
  H_e^{\eta}({\bm k})=\frac{k^2}{2m^*}+[\lambda\eta+A_1k^2\eta+A_2(k_x^3-3k_xk_y^2)]\sigma_z,\label{normal_H}
\end{equation}
in the basis $[c_{\eta {\bm k}\uparrow}, c_{\eta {\bm k}\downarrow}]$, with
$c_{\eta {\bm k} s}$ the annihilation operator for an electron in valley $\eta$ at momentum ${\bm k}$ with spin $s = \uparrow, \downarrow$. 
We obtain this effective Hamiltonian from the ${\bm k}\cdot {\bm p}$ Hamiltonian near the $\pm \bm K$
valleys in Ref.~\onlinecite{Fang2015} by the L\"owdin partition method~\cite{Lowdin1951,Wang_L2014}.
Here, the $x\ (y)$-axis points along the zigzag (armchair) direction as in Fig.~\ref{fig:hex_nodal_points}(a), $m^*$ denotes the effective mass, $\lambda$ and $A_{1,2}$ are SOC strengths, and $\sigma_{x, y, z}$ are the Pauli matrices in spin space.
Material parameters are provided as Supplemental Material \cite{suppl}.

Including superconductivity with $s$-wave pairing and an in-plane magnetic field, the Bogoliubov-de Gennes (BdG) Hamiltonian in the basis $[c_{\eta {\bm k}\uparrow},c_{\eta {\bm
      k}\downarrow},c^{\dagger}_{-\eta-{\bm k}\uparrow},c^{\dagger}_{-\eta-{\bm
    k}\downarrow}]$ is
\begin{eqnarray}
  H_{\rm BdG}^{\eta}({\bm k})
  &=&(\frac{k^2}{2m^*}-\mu)\tau_z+[\lambda\eta+A_1k^2\eta+A_2(k_x^3-3k_x\nonumber\\
    &&\mbox{}\times k_y^2)]\sigma_z+V_x\sigma_x\tau_z+V_y\sigma_y+\Delta\sigma_y\tau_y,
    \label{eq:H_BdG}
\end{eqnarray}
where $\mu$, $\tau_{x, y, z}$, $\Delta$, and $V_{x,y}$ are the chemical
potential, Pauli matrices in particle-hole space, the superconducting gap, and
the Zeeman energy terms due to the magnetic field, respectively.

The BdG Hamiltonian $H_{\rm BdG}^{\eta}({\bm k})$ has a particle-hole symmetry
(PHS) $\mathcal{P}H_{\rm
  BdG}^{\eta}({\bm k}){\mathcal{P}}^{-1}=-H_{\rm BdG}^{-\eta}(-{\bm k})$ where
$\mathcal{P}=\tau_x\mathcal{K}$, with $\mathcal{K}$ being the complex conjugation operator.
Although time-reversal symmetry (TRS) $\mathcal{T}=i\sigma_y\mathcal{K}$ is broken by the magnetic field, $H_{\rm
  BdG}^{\eta}({\bm k})$ respects an effective TRS $\tilde{\mathcal{T}}H_{\rm
  BdG}^{\eta}({\bm k}){\tilde{\mathcal{T}}}^{-1}=H_{\rm BdG}^{-\eta}(-{\bm k})$ where
$\tilde{\mathcal{T}}=M_{xy}\mathcal{T}$, with $M_{xy}=-i\sigma_z\tau_z$ the mirror symmetry in the monolayer plane.
Therefore, $H_{\rm BdG}^{\eta}({\bm k})$ has the chiral symmetry $\mathcal{C}H_{\rm BdG}^{\eta}({\bm k}){\mathcal{C}}^{-1}=-H_{\rm BdG}^{\eta}({\bm k})$ with
$\mathcal{C}=\mathcal{P}\tilde{\mathcal{T}} = \sigma_x\tau_y$.
As a result, $H_{\rm BdG}^{\eta}({\bm k})$ is in class BDI, which is trivial in
two dimensions for gapped systems~\cite{Schnyder2008,Chiu2016}, but can be
nontrivial for nodal systems.
We reduce the dimension to one by fixing two orthogonal directions
${\bm k}_\parallel$ and ${\bm k}_\perp$ in momentum space, and considering each
$H_{\rm BdG}^{\eta}({\bm k}_\perp, {\bm k}_\parallel)$ at a fixed ${\bm
  k}_\parallel$ separately~\cite{Sato2011}.
Although $\mathcal{P}$ and $\tilde{\mathcal{T}}$ are in general not symmetries of the one-dimensional (1D) Hamiltonian $H_{\rm BdG}^{\eta}({\bm k}_\perp, {\bm k}_\parallel)$ at a fixed ${\bm k}_\parallel$ \cite{Varjas2018}, because they flip the sign of both ${\bm k}_\parallel$ and ${\bm k}_\perp$ \footnote{The 1D PHS or TRS that only flip ${\bm k}_\perp$ require an extra unitary symmetry that maps ${\bm k}_\parallel \rightarrow -{\bm k}_\parallel$, 
but we find that no such symmetry exists for generic $({\bm k}_\perp, {\bm k}_\parallel)$.}; $\mathcal{C}$ is a symmetry for any choice of ${\bm k}_\parallel$.
Therefore the 1D Hamiltonians $H_{\rm
  BdG}^{\eta}({\bm k}_\perp, {\bm k}_\parallel)$ at a fixed ${\bm
  k}_\parallel$ belong to class AIII \footnote{For the armchair direction ${\bm k}_\parallel
  = k_y \hat{\bm y}$, although the 1D Hamiltonian
  of the continuum model belongs to class BDI, it turns out to be in class AIII in
  the tight-binding models, due to the absence of a point group symmetry that maps $y \rightarrow -y$.}, and are thus characterized by an integer
topological invariant: the winding number \cite{Schnyder2008}.

\section{Bulk nodal points}
We begin investigating the topological phases of $H_{\rm BdG}^{\eta}({\bm k})$ by finding the gap-closing conditions, which determine the bulk nodal points.
Due to chiral symmetry, $H_{\rm BdG}^{\eta}({\bm k})$ can be brought to a block
off-diagonal form \cite{Beri2010, Sato2011},
with the upper off-diagonal element
\begin{eqnarray}
  A_{\eta}({\bm
    k})&=&-(\frac{k^2}{2m^*}-\mu)+[\lambda\eta+A_1k^2\eta+A_2(k_x^3-3k_xk_y^2)]\sigma_z\nonumber\\
  &&\mbox{}-V_x\sigma_x+V_y\sigma_y+i\Delta\sigma_z.
\end{eqnarray}
The gap-closing condition ${\rm det}[A_{\eta}({\bm k})]=0$ gives rise to two requirements:
\begin{subequations}
\begin{align}
\lambda\eta+A_1k^2\eta+A_2(k_x^3-3k_xk_y^2)&=0, \label{nodal}\\
\mu\pm \sqrt{V_x^2+V_y^2-\Delta^2}&=\frac{k^2}{2m^*}. \label{mom_circ}
\end{align}
\label{eq:gap_closing}
\end{subequations}
\noindent The first is the vanishing of spin splitting due to SOC [see Eq.~(\ref{normal_H})], and the
second is the magnetic field closing the bulk gap at the Fermi circle without SOC.
These two conditions arise because closing the gap with the magnetic field brings together bands that are coupled by SOC.
The bands thus repulse, except at points in momentum space where the SOC vanishes and the gap closes.
Such points manifest as crossings between the spin-split conduction bands in the
normal-state dispersion, which are present near $\pm \bm K$ valleys in monolayer
\ce{MoX2} (X=S, Se, Te) but not \ce{WS2}, due to the relative strengths of SOC
  contributions from the $d$ orbitals on the transition metal atoms and the $p$
  orbitals on the chalcogen atoms~\cite{Liu2013, Kosmider2013,
  Kormanyos2015}. 
Therefore, the requirement~\eqref{nodal} is not met in \ce{WS2}, and we focus on \ce{MoX2} in the following.
The gap-closing requirements \eqref{eq:gap_closing} are independent of the in-plane magnetic field orientation, so we set $V_y  = 0$ in the following.
Solving Eq.~\eqref{nodal} limits $k$ to $k_{c1}\le k\le k_{c2}$ with $k_{c1,c2}=k_0\pm k_0^2/(2A_0)$, $k_0=\sqrt{-\lambda/A_1}$, and $A_0=A_1/A_2$~\cite{suppl}.
Figure~\ref{fig:hex_nodal_points}(b) shows a phase diagram of the gap-closing conditions as a function of $\mu$ and $\sqrt{V_x^2-\Delta^2}$.
The four phase boundaries $\mu=\mu_{1,2}\pm \sqrt{V_x^2-\Delta^2}$ with
$\mu_{1,2}=k^2_{c1,c2}/(2m^*)$ divide the diagram into regimes, with nodal points and therefore possible nontrivial phases in the colored regions (I and II).

\section{Topological phases}
\begin{figure}[!tbh]
\includegraphics[width=0.97\columnwidth]{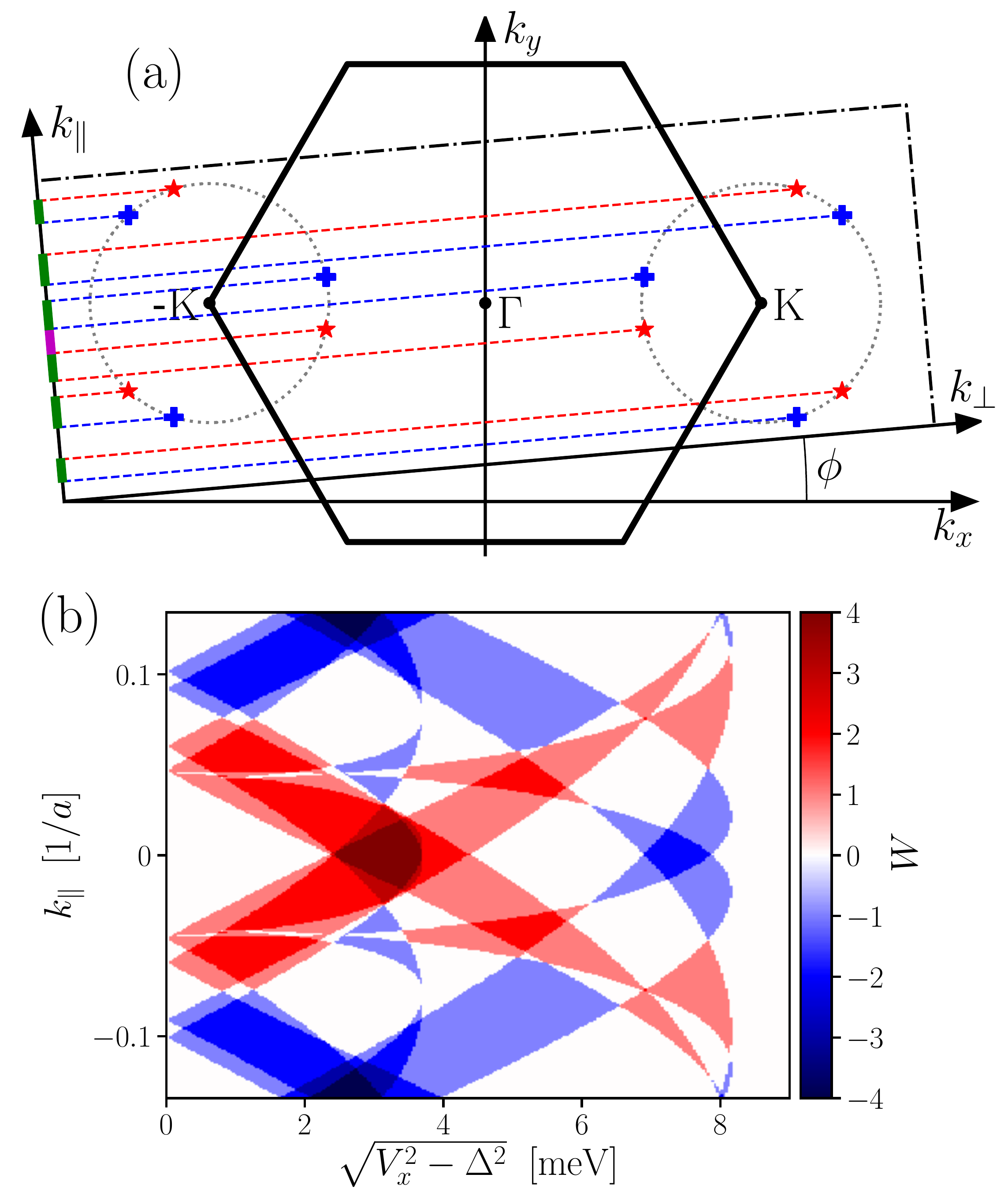}
\caption{(a) Schematic of the hexagonal first Brillouin zone of the monolayer lattice, with nodal points around the high symmetric points $\pm\bm K$.
For arbitrary edge cuts, we deform the Brillouin zone into a rectangle, illustrated by the dash-dotted lines and the $k_\parallel$ and $k_\perp$ axes, and project the nodal points onto $k_\parallel$.
Flat bands of Andreev bound states exist for $k_\parallel$ where the winding number is nonzero (bold colored lines). 
For a generic edge cut, each nodal point generally projects onto a distinct $k_\parallel$, such that the winding number may take various values, {e.g.}~$\pm 1$ (green) or $\pm 2$ (purple) in the sketch.
(b) Phase diagram of the winding number for an edge with $ \phi \approx 1.2^\circ$.
The phase diagram is rich with the winding number $\pm 1$, $\pm 2$, $\pm 3$, or $\pm 4$.
Data is obtained from an 11-orbital tight-binding model for \ce{MoS2} with $\mu=1.8390\ $eV and $\Delta=0.8$ meV, see Supplementary Material.}
\label{fig:other_edge_cut}
\end{figure}

In the gapless regimes of the phase diagram, Fig.~\ref{fig:hex_nodal_points}(c) sketches the nodal points near the $\bm K$ valley along with their chirality $w({\bm k}^i)$.
The chirality of the nodal point at ${\bm k}^i = (k_\perp^i, k_\parallel^i)$ is the winding number around it, and is $\pm 1$ \cite{Beri2010, Sato2011, Schnyder2011}.
The nodal point chirality relates to the winding number $W$ of the
1D Hamiltonian at a fixed $k_\parallel$ through $W(k_\parallel) =\sum_{{k^i_\parallel} < k_\parallel} w({\bm k}^i)$, which means that we can obtain $W(k_\parallel)$ by counting the nodal point projections onto the $k_\parallel$-axis and keeping track of their chirality.
For the zigzag direction ${\bm k}_\parallel = k_x \hat{\bm x}$, the nodal point projections cancel exactly, because the nodal points come in pairs with opposite chirality at each $k_x$, and hence $W(k_x) = 0$ always.
For any other direction, the nodal points do not cancel, and nontrivial phases thus exist for all directions ${\bm k}_\parallel$ other than zigzag.
We show the projections of the nodal points on the armchair direction ${\bm
  k}_\parallel = k_y \hat{\bm y}$, and the corresponding segments of the
$k_y$-axis where $W(k_y) \neq 0$ (solid green lines).
In regime I, there are two momentum circles \eqref{mom_circ} near the $\bm K$ valley, with six nodal points each for a total of twelve.
The nodal points divide the $k_y$-axis into thirteen segments, with six segments topologically nontrivial.
In regime II, there is only one momentum circle with six nodal points, such that
the $k_y$-axis separates into seven parts, with three nontrivial.
At the boundary between regimes I and II (marked as III in the figure), pairs of nodal points of opposite chirality overlap on one momentum circle, such that only the other circle contributes to the winding number $W$, similar to regime II.
The nodal points near the $-\bm K$ valley are symmetric to the ones near $\bm K$ in $k_x$ [see also Fig.~\ref{fig:other_edge_cut}(a)].
The preceding analysis applies equally to all three \ce{MoX2} monolayers.
In the following, we explore further details of the topological phases, focusing on nodal point projections on the armchair direction.
Although we show examples for specific materials, we have verified that the physics is qualitatively the same for all three~\cite{suppl}.

\begin{figure}[!tbh]
\includegraphics[width=0.97\columnwidth]{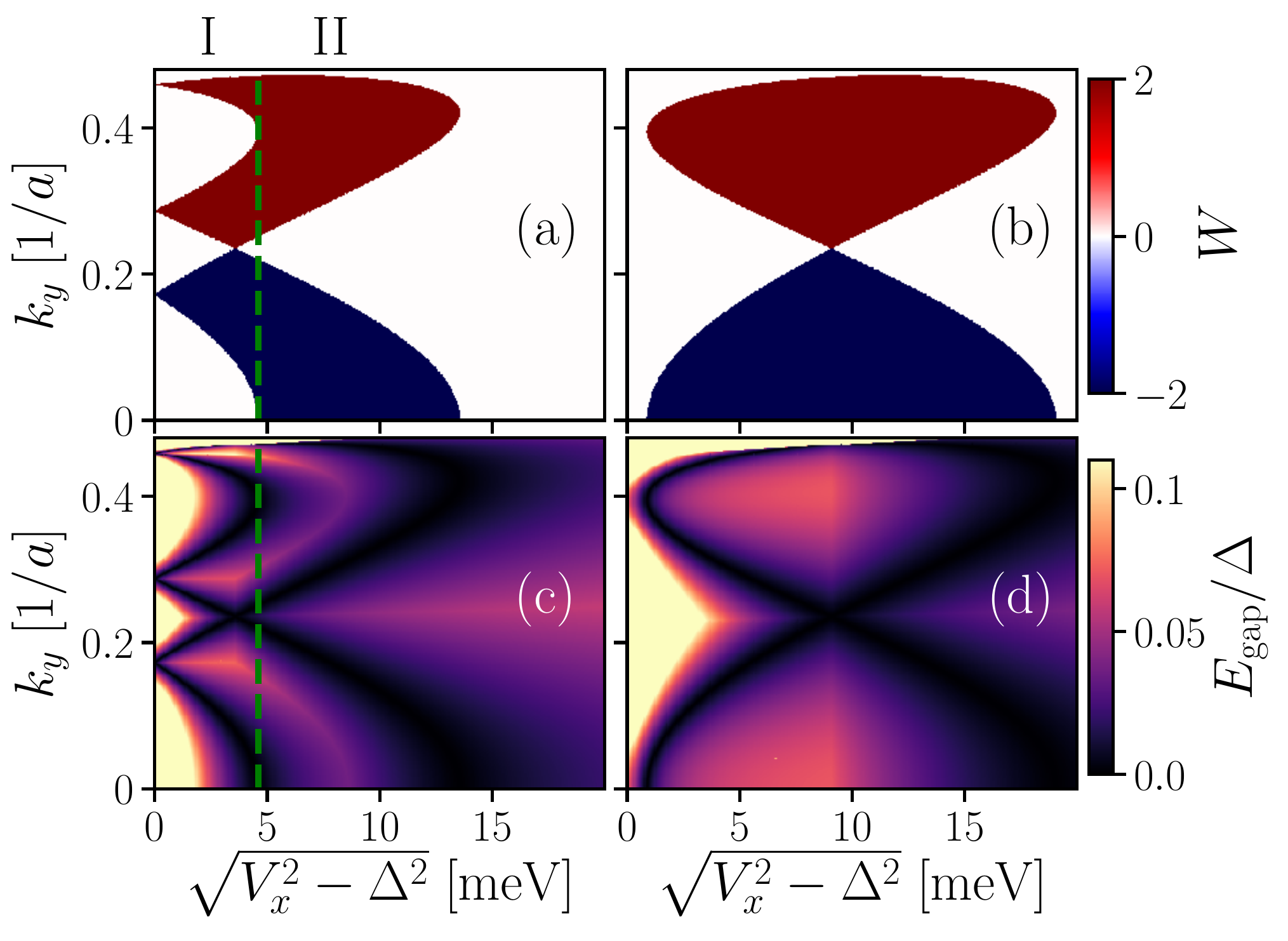}
\caption{Topological phase diagrams for the armchair direction ${\bm k}_\parallel = k_y \hat{\bm y}$ of monolayer \ce{MoSe2}.
The winding number as a function of $k_y$ and $\sqrt{V_x^2-\Delta^2}$ with
(a) $\mu_1<\mu<(\mu_1+\mu_2)/2$ and (b) $\mu<\mu_1$, in regimes I and II of Fig.~\ref{fig:hex_nodal_points}.
The phase diagrams for $(\mu_1+\mu_2)/2<\mu<\mu_2$ and $\mu>\mu_2$ are similar to (a) and (b), respectively, but with opposite winding numbers.
(c)-(d) The corresponding topological excitation gap $E_{\rm gap}$ to (a) and (b) separately.
Data is obtained using the continuum model \eqref{eq:H_BdG}, and $a$ is
the lattice constant of the \ce{MX2} lattice.}
\label{fig:winding_gap}
\end{figure}

To complement the analysis of nodal point projections,
Figs.~\ref{fig:winding_gap}(a) and \ref{fig:winding_gap}(b) show computed phase diagrams of the winding number as a function of $k_y$ and $\sqrt{V_x^2-\Delta^2}$ at two chemical potentials, $\mu_1<\mu<(\mu_1+\mu_2)/2$ in (a) and $\mu<\mu_1$ in (b), respectively, representative of regimes I and II of Fig.~\ref{fig:hex_nodal_points}.
The phase diagrams are even in $k_y$, and the winding number is $\pm 2$ due to equal contributions from the $\pm\bm K$ valleys.
The phase boundaries in Fig.~\ref{fig:hex_nodal_points}(b) determine the range of the nontrivial regions in $\sqrt{V_x^2-\Delta^2}$, while the maximum extent along $k_y$ is bounded from above by $|k_y| \leq k_0$, independent of $\mu$ and $\sqrt{V_x^2 - \Delta^2}$ \cite{suppl}.
Sweeping over $\sqrt{V_x^2-\Delta^2}$ in (a), the phase diagram transitions from regime I to II indicated by the vertical dashed line, such that the number of topologically nontrivial segments along $k_y$ changes from six to three (also counting $-k_y$).
In contrast, (b) is exclusively in regime II.

\begin{figure}[!tbh]
\includegraphics[width=0.97\columnwidth]{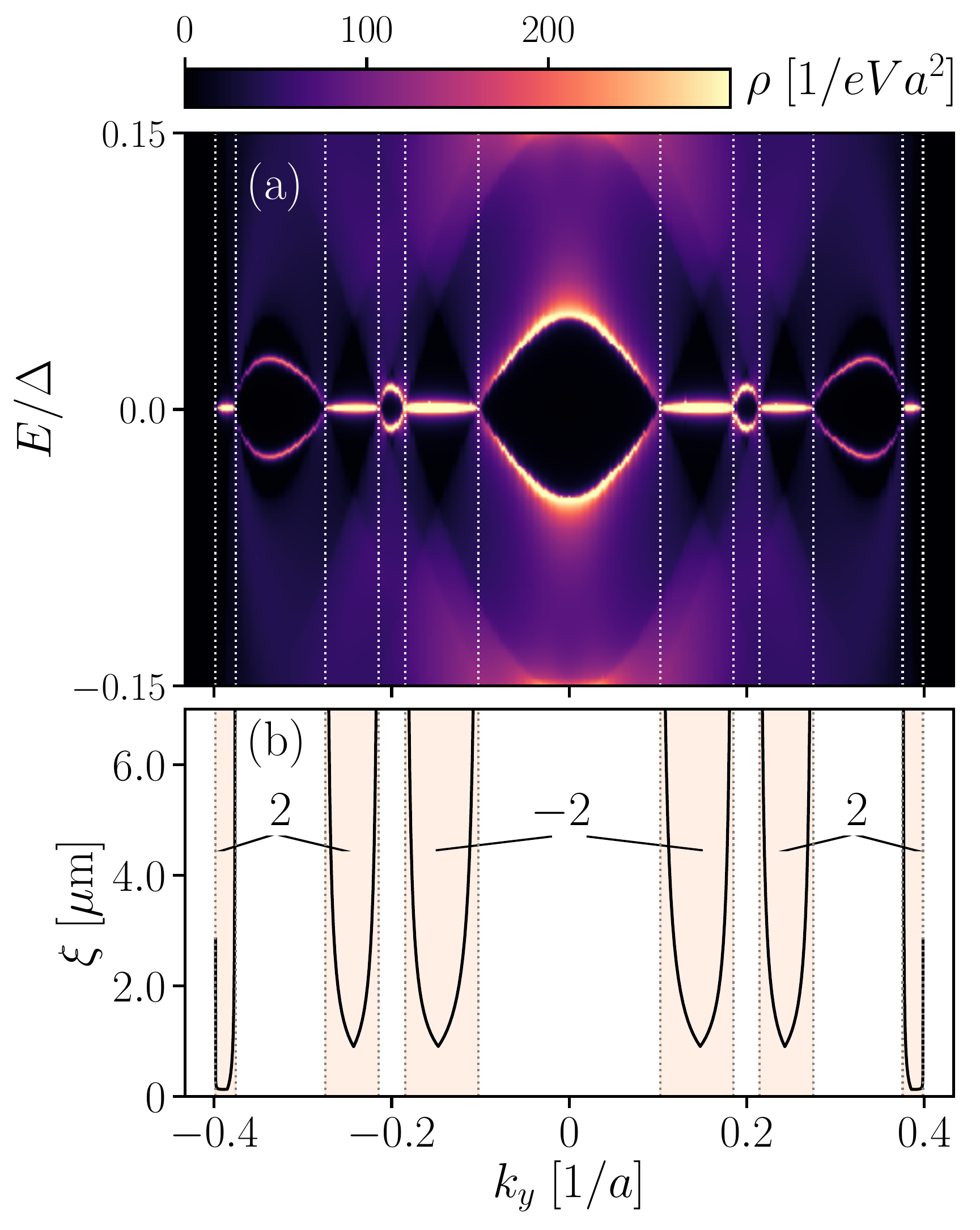}
\caption{(a) Density of states at the armchair edge as a function of $k_y$ for monolayer \ce{MoS2}, with parameters in regime I of Fig.~\ref{fig:hex_nodal_points}(b).
Flat bands of zero-energy Andreev edge states where the winding number is nonzero between nodal point projections.
(b) Decay length of the edge states in the topologically nontrivial phase.
The nontrivial phases are marked by the shaded regions with the nonzero winding numbers in the insets.
Data is obtained using an 11-orbital tight-binding model with $\mu = 1.8337$ eV, $\sqrt{V_x^2 - \Delta^2} = 1.5$ meV, and $\Delta=0.8$ meV, see Supplementary Material.} 
\label{fig:LDOS}
\end{figure}

\section{Excitation gap and edge states}
Topologically nontrivial phases are protected by the topological excitation gap, which we define as $E_{\rm
  gap}({\bm k}_\parallel) = \mathrm{min}_{n, {\bm k}_\perp} |E_{n}({\bm
  k}_\parallel, {\bm k}_\perp)|$, where $E_{n}({\bm k})$ is the spectrum of
$H_{\rm BdG}^\eta({\bm k})$, with $n$ a band index. Figures \ref{fig:winding_gap}(c) and \ref{fig:winding_gap}(d) show maps of the topological excitation gap corresponding to the phase diagrams (a) and (b), respectively.
In the nontrivial phase, we see that $E_{\rm gap} \lesssim 0.1 \Delta$ for \ce{MoSe2}, and similarly find $E_{\rm gap} \lesssim 0.04\Delta$ for \ce{MoS2}, and $E_{\rm gap} \lesssim 0.2\Delta$ for \ce{MoTe2} \cite{suppl}.
Here, we emphasize that $\Delta$ may represent intrinsic superconductivity, which means that no proximity effect is required, and interface effects that tend to reduce the gap further are thus absent.

In a topologically nontrivial phase, edge states manifest at a monolayer lattice termination boundary.
We investigate the edge states at an armchair edge by calculating the local density of states at the boundary, $\rho(E, x_B, k_y)=-\frac{1}{\pi}{\rm Tr}[{\rm Im}G(E, x_B, k_y)]$, with $E$ the energy, $x_B$ the coordinate of the armchair edge, and $G$ the surface Green's function \cite{Datta1995}.
Figure \ref{fig:LDOS}(a) shows the local density of states obtained using parameters from regime I of Fig.~\ref{fig:hex_nodal_points}(b), {i.e.}~with $12$ nodal points per valley.
At zero energy, there are six sections of Andreev flat bands connecting nodal points, which exactly match the topologically nontrivial phases with nonzero winding number, marked by the vertical dotted lines, and the shaded regions in Fig.~\ref{fig:LDOS}(b).
In Fig.~\ref{fig:LDOS}(b), we also present the decay length of the topologically nontrivial edge states, and see that it is of the order \SI{1}{\mu m} here.

\section{Arbitrary edge directions}
Although we have so far focused on an armchair edge, topologically nontrivial regimes exist for all lattice termination edges except zigzag.
Using tight-binding models to simulate the \ce{MX2} lattice [Fig.~\ref{fig:hex_nodal_points}(a)] with Kwant \cite{Groth2014}, we characterize a lattice termination edge with a superlattice vector $\bm T$ at the angle $\phi$ relative to the armchair direction~\cite{suppl}.
To investigate topological phases, we deform the hexagonal first Brillouin zone into the rectangle spanned by primitive reciprocal vectors $\hat{{\bm k}}_\parallel$ and $\hat{{\bm k}}_\perp$, which are parallel and, respectively, transverse to $\bm T$ \cite{Delplace2011}, and project the nodal points onto the $k_\parallel$-axis [Fig.~\ref{fig:other_edge_cut}(a)].
As before, flat bands exist in segments of the $k_\parallel$-axis where the winding number is nonzero.
Unlike an armchair edge, the nodal points near $\pm\bm K$ generally do not project pairwise onto the same $k_\parallel$ at a generic boundary, and the winding number can take other values than $\pm 2$, e.g., $\pm 1$ (green lines).
Figure~\ref{fig:other_edge_cut}(b) is an example of a phase diagram for an edge direction with $\phi \approx 1.2^\circ$, and shows that the winding number can be $\pm 1$, $\pm 3$, and even $\pm 4$.
For generic lattice terminations other than armchair, nodal topological phases are thus not only present, but also manifest in rich phase diagrams with large winding numbers.

\begin{table}[ht]
\caption{Chemical potentials $\mu_{1,2}$ in meV for \ce{MoS2}, \ce{MoSe2}, and \ce{MoTe2} [see also Fig.~\ref{fig:hex_nodal_points}(b)], obtained from the continuum model.}
\begin{tabular}{c | c c c}
\hline\hline
~ & \ce{MoS2} & \ce{MoSe2} & \ce{MoTe2} \\
\hline
$\mu_1$ & $32.6$& $126.7$ & $136.1$ \\
$\mu_2$ & $34.5$ & $143.0$ & $184.5$ \\
\hline\hline
\end{tabular}
\label{tab:kp_mu}
\end{table}

\section{Summary and discussion}
We have shown that a superconducting monolayer \ce{MoX2} (X$=$S, Se, Te) can become a nodal TSC in the presence of an
in-plane magnetic field. The bulk nodal points occur at special momenta near
$\pm\bm K$ valleys in the Brillouin zone where the spin splitting due to SOC vanishes, and can be $6$ or $12$ in each valley.
For all lattice termination edges except zigzag, the edge projections of the
nodal points are connected by flat bands of zero-energy Andreev edge
states. These flat bands are protected by chiral symmetry. Our conclusions are based on a study of both continuum and atomic tight-binding models.

Finally, we address experimental feasibility.
It is possible to produce high-quality monolayer \ce{MoX2} crystals with low impurity
densities, and sizes in the tens of microns or even millimeters
\cite{Coleman2011, Chowalla2013, Wang2014, Chen2017}.
Such large samples may guarantee that the topological Andreev edge states at opposing edges are well separated.
In addition, recent experiments show that thin films even down to monolayers of
\ce{MoX2} become superconducting in the conduction band at carrier densities
$\gtrsim 6\times 10^{13}\ $cm$^{-2}$ \cite{Shi2015, Lu2015, Ye2012}, which
translates to a minimum chemical potential $\mu_0$ for superconductivity of
$153\ $meV (\ce{MoS2}), $120\ $meV (\ce{MoSe2}), and $117\ $meV (\ce{MoTe2}).
The mismatch of $\mu_0$ and $\mu_{1,2}$ in \ce{MoS2} implies that intrinsic
superconductivity is not suitable to realize the nodal topological phase in
\ce{MoS2}, but this can potentially be overcome using the proximity
effect.
In addition, a recent experiment indicates possible intrinsic unconventional pairing in \ce{MoS2} at very large doping \cite{Costanzo2018}.
For monolayer \ce{MoSe2} and \ce{MoTe2}, $\mu_0$ is close to $\mu_{1,2}$ in Fig.~\ref{fig:hex_nodal_points}(b) [see Table~\ref{tab:kp_mu}], and therefore these two materials are promising candidates for realizing nodal TSCs.
For experimental detection, aside from tunneling measurements,
the character of bulk nodal points could be probed using quasiparticle
interference or local pair-breaking measurements \cite{Hanaguri2007, Allan2013,
  Zhou2013}. Because the flat bands manifest as a zero-energy
  density of states peak in the nontrivial parts of
  the phase diagram Fig.~\ref{fig:hex_nodal_points}(b), it is possible to discern them from other edge states \cite{Rostami2016},
  which generally don't stick to zero energy, by tuning the magnetic field
  and/or chemical potential. If the chiral symmetry is broken, the flat bands
  may split from zero energy. Two possible causes are a perpendicular electric
  field due to asymmetric electrostatic gating, and an out-of-plane Zeeman field. The electric
  field can be avoided by chemical doping \cite{Ye2012,Lu2015} and it is
  possible to align the magnetic field along the in-plane direction to a precision of $\lesssim 0.02^\circ$, such that the out-of-plane projection is
  negligible \cite{Saito2016}.

\acknowledgements
We thank Anton Akhmerov, Alexander Lau, and Valla Fatemi for fruitful discussions. This work was supported by ERC Starting Grant 638760, the Netherlands Organisation for Scientific Research (NWO/OCW), as part of the Frontiers of Nanoscience program,
and the U.S. Office of Naval Research.

L.W.~conceived and initiated the
project. L.W.~contributed to most of the continuum model calculations, T.O.R.~performed most
of the calculations with the $11$-orbital tight-binding model, and
D.S.~performed most of the calculations with the three-orbital tight-binding
model, and some with the continuum model. All authors contributed to writing the paper. 

\bibliographystyle{apsrev4-1}
\bibliography{bibl}

\clearpage
\setcounter{figure}{0}
\setcounter{section}{0}
\onecolumngrid

\section{Supplementary material}

\subsection{Vanishing of spin splitting due to spin-orbit coupling in continuum model for monolayer $\text{MoX}_2$}
The condition for vanishing of spin splitting due to spin-orbit coupling (SOC) in the continuum model is given by Eq.~\eqref{nodal} of the main text:
\begin{equation}
\lambda\eta+A_1k^2\eta+A_2(k_x^3-3k_xk_y^2)=0. \label{nodal2}\\
\end{equation}
With $\eta=1$, we perform an analytic derivation near $\bm K$ valley. By defining
$k_0=\sqrt{-\lambda/A_1}$ and $A_0=A_1/A_2$, Eq.~(\ref{nodal2}) becomes
\begin{equation}
  k_x^3+A_0k_x^2-3k_xk_y^2+A_0(k_y^2-k_0^2)=0,
\end{equation}
which is a cubic equation in $k_x$ for a fixed $k_y$. By considering that
$|k_y|$ and $k_0$ are much smaller than $|A_0|$, this equation has three real
solutions approximately, $k_{x1,x2}=\frac{4k_y^2-k_0^2}{2A_0}\pm
\sqrt{k_0^2-k_y^2}$ and $k_{x3}=-A_0$ for $|k_y|\le k_0$ whereas only one real
solution approximately, $k_x=-A_0$ for $|k_y|>k_0$. Note that the solution
$k_x=-A_0$ is unphysically large in the continuum model, and we therefore reject it. As a result, the cubic equation has two physical solutions $k_{x1,x2}$
only when $|k_y|\le k_0$, and no solutions otherwise. This gives rise to an upper limit of $|k_y|$ in the
topologically nontrivial phase for armchair direction.

The magnitude of the momentum where the spin splitting due to SOC vanishes is calculated by
$k^2=k_{x1,x2}^2+k_y^2\approx
\pm\frac{\sqrt{k_0^2-k_y^2}}{A_0}[4(k_0^2-k_y^2)-3k_0^2]+k_0^2$. Due to
$|k_y|\le k_0$, we have the maximum of $k$,
$k_{c2}=\sqrt{k_0^2-k_0^3/A_0}\approx k_0-k_0^2/(2A_0)$ and the minimum of $k$,
$k_{c1}=\sqrt{k_0^2+k_0^3/A_0}\approx k_0+k_0^2/(2A_0)$.    

Note that the results near $-\bm K$ valley can be easily obtained by replacing $k_x$
by $-k_x$.  

\begin{table}[ht]
\caption{Material parameters for the continuum model of the monolayer
  \ce{MoX2} family [see Eq.~\eqref{normal_H} of the main text]. They are
  obtained from $\bm k \cdot \bm p$ Hamiltonians near $\pm {\bm K}$ valleys in Ref.~\onlinecite{Fang2015} (\ce{MoS2} and \ce{MoSe2}) and in Ref.~\onlinecite{Kormanyos2015}
  (\ce{MoTe2}) by the L\"owdin partition method \cite{Wang_L2014,Lowdin1951}. Here, $a$ is the lattice constant of the monolayer lattice.}
\begin{tabular}{c | c c c c c}
\hline\hline
~ & $a$ [$\AA$] ~ & $m^*/m_0$ ~ & $\lambda$ [\SI{}{meV}] ~ & $A_1/a^2$ [\SI{}{meV}] ~ & $A_2/a^3$ [\SI{}{meV}]\\
\hline
\ce{MoS2} & 3.18 & 0.47 & -1.5 & 36 & -4.88 \\
\ce{MoSe2} & 3.32 & 0.60 & -10.6 & 45.5 & -6.23 \\
\ce{MoTe2} & 3.516 & 0.62 & 18 & -56.33 & 15.13 \\
\hline\hline
\end{tabular}
\label{tab:kp_material}
\end{table}

\subsection{Tight-binding Hamiltonians for monolayer $\text{MoX}_2$}
In this section, we introduce the crystal structure symmetries of the monolayer \ce{MoX2} lattice, and the monolayer tight-binding Hamiltonians that we have adopted in our analysis.
The monolayer \ce{MoX2} crystal consists of a two-dimensional triangular lattice
of Mo atoms, sandwiched between two equidistant layers of X atoms X$_A$ above
and X$_B$ below, each also forming a triangular lattice, such that the X$_A$ and
X$_B$ atoms project onto the same position in the layer of Mo atoms [see
  Fig.~\ref{fig:hex_nodal_points}(a) of the main text].
The primitive Bravais lattice vectors are
\begin{equation}
\boldsymbol{a}_1 = a \hat{\bm x}, \quad
\boldsymbol{a}_2 = a( \frac{1}{2}\hat{\bm x} + \frac{\sqrt{3}}{2}\hat{\bm y}).
\label{eq:lat_vecs}
\end{equation}
The Mo atoms lie in the $xy$-plane, with the planes of X$_A$ and X$_B$ atoms above and below at distances $\pm d/2$, such that $d$ is the separation between the two planes of X atoms.
The monolayer has threefold rotational symmetry, a mirror symmetry in the $yz$-plane $M_{yz}: x \rightarrow -x$, and the $xy$ mirror symmetry $M_{xy}: z \rightarrow -z$.

\subsubsection{The three-orbital model} \label{section:3orb}
In Ref.~\cite{Liu2013} a simple tight-binding model was proposed to reproduce
the low-energy spectrum near the $\pm\bm K$ point where only $d_0\equiv d_{z^2},
d_{xy}$, and $d_{x^2-y^2}$ orbitals of the Mo atom are considered due to them having the largest contribution to the electronic band.
The detailed structure of the spinless three-orbital Hamiltonian $H_0$ in the basis $(d_{0}, d_{xy}, d_{x^2-y^2})$ in Bloch form reads
\begin{equation}
H_0=\pmat{h_0 & h_1 & h_2\\
h_1^* & h_{11} & h_{12} \\
h_2^* & h_{12}^* & h_{22}
},
\end{equation}
with the following matrix elements for nearest-neighbor hopping:
\begin{eqnarray}
h_0 &=& 2t_0 (2\cos \xi\cos\gamma+\cos2 \xi)+\varepsilon_1
-\mu
\notag\\
h_1 &=&2 i t_1 (\sin 2\xi +\sin\xi\cos\gamma)
-2\sqrt{3}t_2\sin\xi\sin\gamma,\notag\\
h_2&=&2 i \sqrt3 t_1 \cos\xi\sin\gamma +
2t_2(\cos 2\xi-\cos\xi\cos\gamma),\\
h_{11}&=&t_{11} (\cos\xi\cos\gamma+2 \cos2 \xi)
+3 t_{22} \cos\xi \cos\gamma+\varepsilon_2-\mu,\notag\\
h_{22}&=&3t_{11} \cos\xi\cos\gamma
+t_{22}(\cos\xi\cos\gamma +2\cos2\xi )+\varepsilon_2-\mu,
\notag\\
h_{12}&=&\sqrt3 (t_{22}-t_{11})\sin\xi\sin\gamma
+4 i t_{12} \sin\xi(\cos\xi-\cos\gamma),\notag
\end{eqnarray}
where $\xi=k_xa/2$ and $\gamma=\sqrt 3 k_ya/2$.

Due to the heavy Mo atoms, there is a large SOC of strength $\lambda$, such that the spinful Hamiltonian reads
\begin{equation}\label{3band}
H_1 =\sigma_0 \otimes H_0 
+ \frac{\lambda}{2} \sigma_z \otimes L_z
\end{equation}
with $\sigma$ the Pauli matrices acting on the spin degree of freedom, and $(L_z)_{kl}=2i\varepsilon_{1kl}$, the matrix elements of $L_z$ in the basis $(d_0,d_{xy},d_{x^2-y^2})$, with $\epsilon$ the Levi-Civita symbol.
The starting tight-binding parameters are displayed in Table~\ref{tab:3bnd_par}, obtained from the nearest-neighbor model in Ref.~\cite{Liu2013}.
In addition we consider a chemical potential $\mu$ tuned into the lowest conduction band.

\begin{table}[ht]
\caption{Tight-binding parameters for the family of \ce{MoX2} materials in eV from a generalized-gradient approximation fit of first-principle data~\cite{Liu2013}.}
\begin{tabular}{c | c c c c c c c c c}
\hline\hline
~ & $\varepsilon_1$ ~ & $\varepsilon_2$ ~ & $t_0$ ~ & $t_1$ ~ & $t_2$ ~ & $t_{11}$ ~ & $t_{12}$ ~ & $t_{22}$ ~ & $\lambda$ \\
\hline
\ce{MoS2}  & 1.046 ~ & 2.104 ~ & $-0.184$ ~ & 0.401 ~ & 0.507 ~ & 0.218 ~ & 0.338 ~ & 0.057 ~ & 0.073 ~\\
\ce{MoSe2} & 0.919 ~ & 2.065 ~ & $-0.188$ ~ & 0.317 ~ & 0.456 ~ & 0.211 ~ & 0.290 ~ & 0.130 ~ & 0.091 ~\\
\ce{MoTe2} & 0.605 ~ & 1.972 ~ & $-0.169$ ~ & 0.228 ~ & 0.390 ~ & 0.390 ~ & 0.207 ~ & 0.239 ~ & 0.107 ~\\
\hline\hline
\end{tabular}
\label{tab:3bnd_par}
\end{table}

As observed in Ref.~\cite{Liu2013} from first-principle calculations, the
\ce{MoX2} materials have crossings between the spin-split conduction bands near
$\pm \bm K$ valleys.
As mentioned in the main text, these crossings are crucial to realize
topological nodal superconductivity, because the spin splitting due to SOC vanishes at the crossing points.
Unfortunately, the three-orbital model in Eq.~\eqref{3band} (even if extended to next-nearest hopping) does not reproduce the expected spin-orbit crossings of the conductance bands.
This is due to the model not properly including the effect of $p$ orbitals from the X atoms.
Extended models with a larger basis containing orbitals from X atoms correctly reproduce the crossings present in the first-principle data~\cite{Fang2015}.

To solve this problem, we extend here the three-orbital model to include the effects of $p_x$ and $p_y$ orbitals near the $\pm \bm K$ points by renormalizing the tight-binding parameters to include virtual hoppings to these orbitals.
Since we are interested only in physics near the $\pm\bm K$ points, where the physics is largely dominated by $d_{0}$, we neglect the renormalization of the other orbitals.
Thus, the hopping integrals between $d_0$ orbitals are renormalized by virtual hoppings on the $p_{x,y}$ orbitals from the X atoms.

The wave functions for the spinless $p_x$ and $p_y$ orbitals read
\begin{equation}
p_x=-\frac{1}{\sqrt 2}(|1,1\rangle - |1,-1\rangle),
\quad
p_y=\frac{i}{\sqrt 2}(|1,1\rangle +|1,-1\rangle),
\end{equation}
in terms of the orbital angular momentum eigenstates.
Therefore 
\begin{equation}
\avg{p_y|\bm L\cdot \bm S|p_x}
=\avg{p_y|\big[\frac{1}{2}(L^+S^-+L^-S^+)+L_z S_z\big]|p_x},
\end{equation}
with $\bm L$ and $\bm S$ the vectors of orbital and spin angular momentum operators, respectively, and $\mathcal{O}^{\pm} = \mathcal{O}_x \pm i \mathcal{O}_y$, $\mathcal{O} \in \{L, S \}$.
We keep only the energetically most important contribution due to the spin-conserving term
\begin{equation}
\avg{p_y|L_z S_z|p_x}=\frac{i}{2}\sigma_z.
\end{equation}
The spin-flip terms are energetically more expensive, involving transitions to higher energy states, and may be neglected to a first order approximation~\cite{Roldan2014}.

The tight-binding equations for the $j$th unit cell of \ce{MoX2} read:
\begin{eqnarray}\label{pd_coupling}
(E-\e_p)c_{x,j}&=&\sum_k t_{x0}(\bm n_k)g_k
-\frac i2 \sigma_z c_{y,j}\\
(E-\e_p)c_{y,j}&=&\sum_k t_{y0}(\bm n_k)g_k
+\frac i2 \sigma_z c_{x,j}\\
(E-\e_d) g_j &=&\sum_k [t_{0x}(\bm n_k)c_{x,k}
+t_{0y}(\bm n_k)c_{y,k}],
\end{eqnarray}
with the sums running over all cells $k$ available through nearest-neighbor hopping.
We denote with $c_{x/y, k}$ the amplitude of an electron in orbital $p_x$ or $p_y$ in cell $k$, and $g_k$ the amplitude of an electron in the $d_{0}$ orbital in cell $k$.
The onsite energy for being in the $p_{x/y}$ or $d_0$ orbitals is $\varepsilon_p$ and, respectively, $\varepsilon_d$. 
The nearest-neighbor hopping from orbital $\beta$ to $\alpha$ from the current cell $j$ to nearby cell $k$ is denoted by $t_{\alpha\beta}(\bm n_k)$, with orbitals $\alpha,\beta \in \{0\equiv d_0 ,x\equiv p_x,y\equiv p_y\}$.
The unit vector $\bm n_k$ points in the hopping direction along the bond.

We solve the equations at energy $E$ close to the conduction band minimum at $\pm\bm K$.
Eliminating the equations involving the $p$ orbitals, we obtain the renormalization of the hopping integrals between $d_0$ orbitals:
\begin{subequations}
\begin{eqnarray}\label{new_soc}
(E-\e_d)g_j&=&\sum_{kl}\bigg\{\big[t_{0x}(\bm n_k)t_{x0}(\bm n_l)+
t_{0y}(\bm n_k)t_{y0}(\bm n_l)
\big]\label{e_corr}\\
&&-\frac{i\sigma_z}{2(E-\e_p)}\big[
t_{0x}(\bm n_k)t_{y0}(\bm n_l)-
t_{0y}(\bm n_k)t_{x0}(\bm n_l)
\big]\bigg\}F(E) g_l,\notag\\
F(E)&=&
\bigg[E-\e_p-\frac{1}{4(E-\e_p)}\bigg]^{-1}.
\end{eqnarray}
\end{subequations}
The first term in Eq.~\eqref{e_corr} is just a tuning of the existing parameters in the model.
Most importantly, the second term in Eq.~\eqref{e_corr} complements the three-orbital model with a spin-obit term which is qualitatively different from the current model.
The structure of the new term recalls the Kane-Mele spin-orbit term in graphene~\cite{Kane2005}.
In graphene, such a spin-orbit term is produced for next-nearest neighbor hopping between $p_z$ orbitals due to virtual transitions to nearest neighbor $p_z$ orbitals.

The hopping terms $t_{0x}$ and $t_{0y}$ depend on $V_{pd\sigma}$, the LCAO (linear combination of atomic orbitals) two-center integrals for $\sigma$ bonds.
They are determined for the lattice orientation in
Fig.~\ref{fig:hex_nodal_points}(a) of the main text by the direction cosines $(l,m,0)$, with the aid of Koster-Slater tables~\cite{Slater1954}:
\begin{equation}
t_{0x}=-\frac{l}{2}V_{pd\sigma}\quad
t_{0y}=-\frac m 2 V_{pd\sigma}.
\end{equation}
Therefore the spin-orbit interaction contribution due to virtual hopping on $p$ orbitals may be abbreviated from Eq.~\eqref{new_soc} to:
\begin{equation}
i\nu_{ij}\beta_\textrm{so}\sigma_z,
\end{equation}
with $\nu_{ij}=\pm 1$, depending whether the hopping between two Mo atoms passes the closest X atom to the right or, respectively, to the left.
The amplitude for the interaction at the conduction band bottom $\varepsilon_c$ reads:
\begin{equation}\label{betaso}
\beta_\textrm{so}=\frac{\sqrt 3}{16}
\frac{F(\varepsilon_c)}
{(\varepsilon_c-\varepsilon_p)
(\varepsilon_c-\varepsilon_d)}.
\end{equation}
Going back to momentum space, we find the final Bloch Hamiltonian:
\begin{equation}
H = H_1+2\beta_{so}[\sin(2\xi)-2\sin(\xi)\cos(\gamma)]
\sigma_z\otimes L_{d_0} +
V_x  \sigma_x \otimes I_{3} + V_y  \sigma_y \otimes I_{3},
\label{eq:ham_3}
\end{equation}
where we have also included Zeeman energy terms $V_x$ and $V_y$ due to an in-plane magnetic field, with $I_3$ the $3\times 3$ identity.
The orbital momentum matrix $L_{d_0}=\textrm{diag}(1,0,0)$ ensures that only the hopping term between $d_0$ orbitals is renormalized by virtual hopping to $p_x$ and $p_y$ orbitals of X atoms.
In fact virtual hoppings will also contribute to renormalize all $d$ orbitals.
Since we are interested in the low-energy physics at $\pm\bm K$ points close to the bottom of conduction band, we neglect further effects.

The result~\eqref{betaso} for the coupling strength overestimates the strength of spin-orbit interaction required to obtain the observed conduction-band spin-orbit splitting.
Instead we perform an additional fit to obtain $\beta_\textrm{so}$.
The lowest conduction bands in the tight-binding model are fitted at $\pm\bm K$
to the continuum model in the main text. The value extracted from the fits and used in our simulations for \ce{MoS2} is $\beta_\textrm{so}\approx\SI{.35}{meV}$,
which reproduces the crossings of the spin-split lowest conduction band. Similarly, the coupling for \ce{MoSe2} and \ce{MoTe2} are approximately $\SI{2.04}{meV}$ and, respectively, $-\SI{3.46}{meV}$.

\subsubsection{The $11$-orbital model}
To further verify our conclusions, we adopt the \emph{ab initio} tight-binding
Hamiltonian for \ce{MoX2} developed by Fang {\it et al.}\ in Ref.~\onlinecite{Fang2015}.
In this tight-binding model, the lowest conduction and topmost valence bands are in good agreement with the band structure obtained from first-principles calculations over the entire Brillouin zone.
Unlike the three-orbital model, which only includes electron orbitals on the transition metal (Mo) atom, this tight-binding model includes orbitals on both the transition metal and chalcogen (X) atoms.
Therefore, the model captures the real three-layer structure of the \ce{MoX2} unit cell, making it possible to investigate effects that rely on the position of individual atoms, such as the orbital effects of a magnetic field.
Furthermore, spin-orbit interaction is naturally included in this model by means of atomic SOC, which reproduces the expected crossings of the spin-split lowest conduction band.

The spinless tight-binding model includes five $d$ orbitals on the Mo atom and six $p$ orbitals on the X atoms per primitive unit cell. 
In Ref.~\onlinecite{Fang2015}, the basis of tight-binding orbitals is chosen to embody the mirror symmetry $M_{xy}$ by forming linear combinations of the $p$ orbitals from X$_A$ and X$_B$ atoms that are eigenstates of $M_{xy}$, effectively treating stacked X$_A$ and X$_B$ as a single composite atom.
To recover a basis that reflects the three-layer structure of the \ce{MoX2}, we disentangle this symmetric basis into its constituent atomic orbitals with a unitary transformation, and use the atomic tight-binding basis
\begin{equation}
\begin{split}
\psi = [ & d_{xz}, d_{yz}, d_{z^2}, d_{xy}, d_{x^2-y^2}, \\
	 &  p_z^A, p_x^A, p_y^A, p_z^B, p_x^B, p_y^B ]^T \\
	 = ~ & [\psi_M, \psi_{X_A}, \psi_{X_B}]^T,
\end{split}
\label{eq:11_basis}
\end{equation}
where in the last line we have grouped the orbitals into vectors by atom.

The tight-binding Hamiltonian includes diagonal atomic onsite terms, and hoppings between various neighboring atoms.
In terms of the atomic blocks \eqref{eq:11_basis}, the spinless Bloch Hamiltonian is given by
\begin{equation}
H_0(\boldsymbol{k}) =
\begin{bmatrix}
H_{\mathrm{Mo} \mathrm{Mo}} & H_{\mathrm{X}_A \mathrm{Mo}}^\dagger & H_{\mathrm{X}_B \mathrm{Mo}}^\dagger \\
H_{\mathrm{X}_A \mathrm{Mo}} & H_{\mathrm{X}_A \mathrm{X}_A} & H_{\mathrm{X}_B \mathrm{X}_A}^\dagger \\
H_{\mathrm{X}_B \mathrm{Mo}} & H_{\mathrm{X}_B \mathrm{X}_A} & H_{\mathrm{X}_B \mathrm{X}_B}
\end{bmatrix}.
\end{equation}
On the diagonal are blocks that consist of onsite terms $h_\alpha$ and intralayer hoppings between nearest neighbor atoms of the same type.
These blocks are given by
\begin{equation}
H_{\alpha \alpha}(\boldsymbol{k}) = h_\alpha + \sum\limits_{j \in \left\{1, 2, 3 \right\}} \left[ T_{\alpha \alpha}^{(j)} e^{-i \boldsymbol{k} \cdot \boldsymbol{\delta}_j} + \mathrm{h.c.} \right],
\end{equation}
where $\alpha \in \left\{\mathrm{Mo}, \mathrm{X}_A, \mathrm{X}_B\right\}$. Here, $T_{\alpha \beta}^{(j)}$ is the matrix of hopping amplitudes from atom $\beta$ to atom $\alpha$ along $\boldsymbol{\delta}_j$ (see Table \ref{tab:hop_dir}).
Off the diagonal, hoppings from Mo atoms to nearest neighbor and next-nearest neighbor X atoms contribute terms of the form
\begin{equation}
H_{\alpha \mathrm{Mo}}(\boldsymbol{k}) = \sum\limits_{j \in \left\{4, 5, 6, 7 \right\}} T_{\alpha \mathrm{Mo}}^{(j)} e^{i \boldsymbol{k} \cdot \boldsymbol{\delta}_j},
\end{equation}
with $\alpha \in \left\{\mathrm{X}_A, \mathrm{X}_B\right\}$.
Finally, interlayer hoppings between X atoms are given by
\begin{equation}
\begin{split}
& H_{\mathrm{X}_B \mathrm{X}_A}(\boldsymbol{k}) = T_{\mathrm{X}_B \mathrm{X}_A}^{(0)} + \\
&\sum\limits_{j \in \left\{1, 2, 3 \right\}} \left[ T_{\mathrm{X}_B \mathrm{X}_A}^{(j)} e^{-i \boldsymbol{k} \cdot \boldsymbol{\delta}_j} + \right. \left. \left(T_{\mathrm{X}_A \mathrm{X}_B}^{(j)}\right)^\dagger e^{i \boldsymbol{k} \cdot \boldsymbol{\delta}_j} \right].
\end{split}
\end{equation}
All the matrices $h_\alpha$ and $T_{\beta \gamma}^{(j)}$ for \ce{MoS2} are provided as supplementary material to the manuscript.
Note that tight-binding parameters for \ce{MoSe2} are also available in Ref.~\onlinecite{Fang2015}, but not for \ce{MoTe2}.
\begin{table}[ht]
\caption{Hopping vectors $\boldsymbol{\delta_j} = \alpha_1 \boldsymbol{a}_1 + \alpha_2 \boldsymbol{a}_2$ in the tight-binding Hamiltonian, with the lattice vectors given in \eqref{eq:lat_vecs}. Adapted from Ref.~\onlinecite{Fang2015}.}
\begin{tabular}{c | c c c c c c c c c}
\hline\hline
 & $\boldsymbol{\delta}_1$ & $\boldsymbol{\delta}_2$ & $\boldsymbol{\delta}_3$ & $\boldsymbol{\delta}_4$ & $\boldsymbol{\delta}_5$ & $\boldsymbol{\delta}_6$ & $\boldsymbol{\delta}_7$ & $\boldsymbol{\delta}_8$ & $\boldsymbol{\delta}_9$ \\
\hline
$\alpha_1$ & $1$ & $0$ & $-1$ & $1/3$ & $-1/3$ & $2/3$ & $2/3$ & $2/3$ & $-4/3$ \\
$\alpha_2$ & $0$ & $1$ & $1$ & $-1$ & $2/3$ & $-1/3$ & $-4/3$ & $2/3$ & $2/3$ \\
\hline\hline
\end{tabular}
\label{tab:hop_dir}
\end{table}

\begin{figure}[!tbh]
\includegraphics[width=0.5\columnwidth]{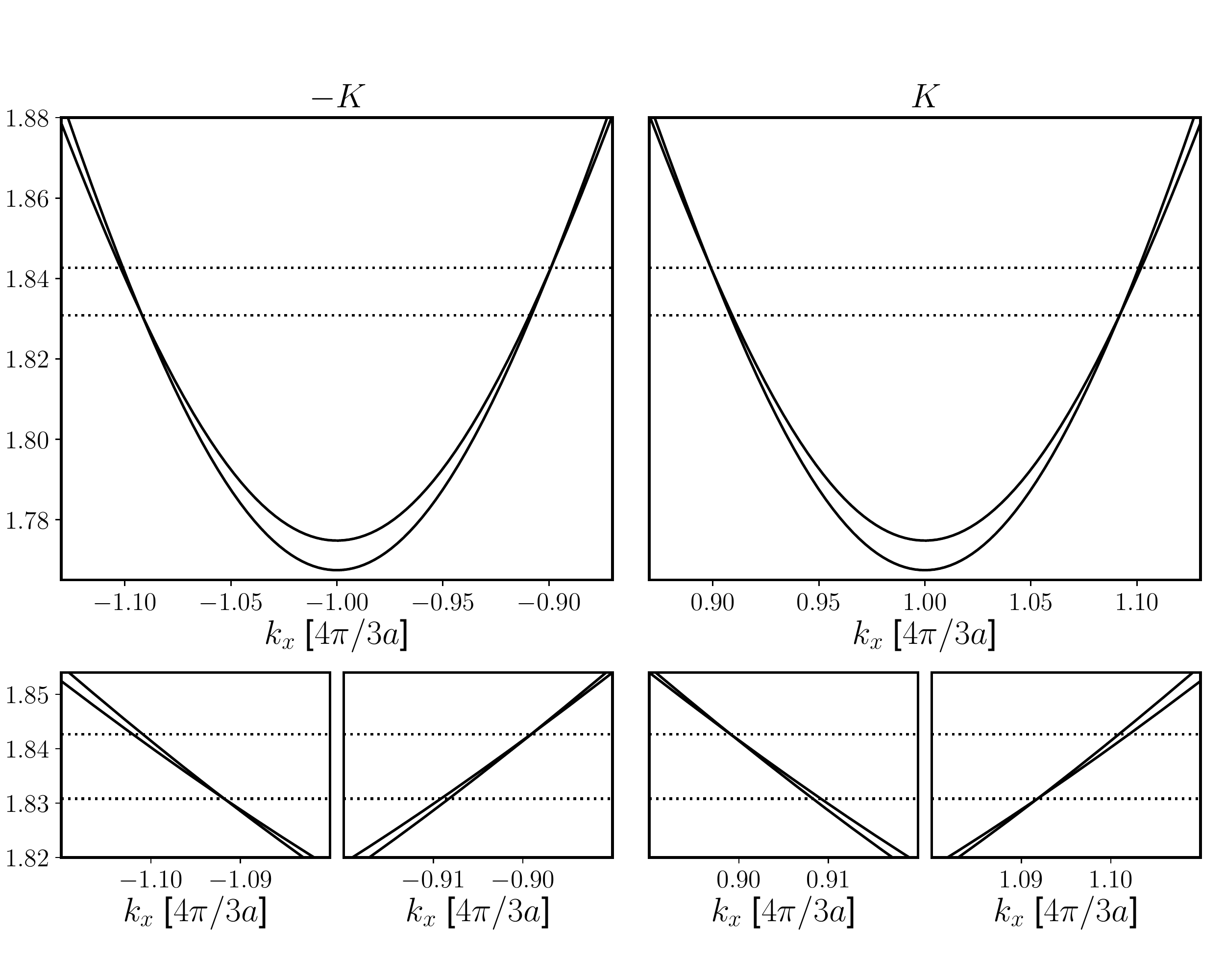}
\caption{Band structure of the normal-state spin-split lowest conduction band in monolayer \ce{MoS2} near the $\pm \bm K$ valleys along the line $k_y = 0$.
The bands are split by spin-orbit interaction, but cross at finite $k_x$ at
points in momentum space where the spin splitting due to SOC vanishes, at energies indicated by the horizontal dotted lines.
The bottom panels show a zoom in on the crossings in each valley.
Such conduction band crossings near $\pm \bm K$ are present in the \ce{MoX2} family of materials.}
\label{fig:crossings_11}
\end{figure}

Spin-orbit coupling is incorporated in the tight-binding model by adding the
atomic spin-orbit interaction terms $\lambda_\alpha \boldsymbol{L} \cdot
\boldsymbol{S}$, with $\lambda_\alpha$ the strength of spin-orbit interaction for atom $\alpha \in \{\mathrm{Mo}, \mathrm{X}_A, \mathrm{X}_B \}$.
Although the lowest conduction band at $\pm \bm K$ is dominated by the $d$ orbitals of Mo atoms, the contribution of X atoms to SOC is necessary to produce the crossings between the lowest spin-split conduction band at $\pm \bm K$ (see Fig.~\ref{fig:crossings_11}), which are essential to realize nodal topological superconductivity.
Including spin, the Bloch Hamiltonian in the basis $[\psi_\mathrm{Mo}^\uparrow, \psi_{\mathrm{X}_A}^\uparrow, \psi_{\mathrm{X}_B}^\uparrow, \psi_\mathrm{Mo}^\downarrow, \psi_{\mathrm{X}_A}^\downarrow, \psi_{\mathrm{X}_B}^\downarrow]^T$ in the presence of an in-plane magnetic field is given by
\begin{equation}
H(\boldsymbol{k}) = \sigma_0 \otimes H_0(\boldsymbol{k}) + H_{\mathrm{SOI}} + V_x  \sigma_x \otimes I_{11} + V_y  \sigma_y \otimes I_{11} \label{eq:H_11},
\end{equation}
with $I_{N}$ the $N \times N$ identity matrix. The spin-orbit interaction matrix $H_{\mathrm{SOI}}$ has the block structure $h_{\alpha \beta}$ with $\alpha, \beta \in \{ \mathrm{Mo}, \mathrm{X}_A, \mathrm{X}_B \}$, and nonzero blocks only for $\alpha = \beta$. Furthermore, $h_{\mathrm{X}_A \mathrm{X}_A} = h_{\mathrm{X}_B \mathrm{X}_B}$ since the atoms are identical.
For SOC on the X atoms, the nonzero matrix elements of $h_{\mathrm{X}_\alpha \mathrm{X}_\alpha}$ with $\alpha = A, B$ are
\[ \begin{split} &\langle p_x^\alpha {\downarrow} | H_{\mathrm{SOI}} |  p_z^\alpha {\uparrow} \rangle = - \lambda_{\mathrm{X}_\alpha}/2, ~
\langle p_y^\alpha {\downarrow} | H_{\mathrm{SOI}} |  p_z^\alpha {\uparrow} \rangle = - i \lambda_{\mathrm{X}_\alpha}/2, ~
\langle p_y^\alpha {\uparrow} | H_{\mathrm{SOI}} |  p_x^\alpha {\uparrow} \rangle = i \lambda_{\mathrm{X}_\alpha}/2, \\
&\langle p_z^\alpha {\downarrow} | H_{\mathrm{SOI}} |  p_x^\alpha {\uparrow} \rangle = \lambda_{\mathrm{X}_\alpha}/2, ~
\langle p_z^\alpha {\downarrow} | H_{\mathrm{SOI}} |  p_y^\alpha {\uparrow} \rangle = i \lambda_{\mathrm{X}_\alpha}/2, ~
\langle p_y^\alpha {\downarrow} | H_{\mathrm{SOI}} |  p_x^\alpha {\downarrow} \rangle = -i \lambda_{\mathrm{X}_\alpha}/2, \end{split} \]
along with their Hermitian conjugates.
Similarly, the nonzero spin-orbit matrix elements of $h_{\mathrm{Mo} \mathrm{Mo}}$ for the Mo atom are
\[\begin{split}
& \langle d_{yz} {\uparrow} | H_{\mathrm{SOI}} |  d_{xz} {\uparrow} \rangle = i \lambda_{\mathrm{Mo}}/2, ~
\langle d_{z^2} {\downarrow} | H_{\mathrm{SOI}} |  d_{xz} {\uparrow} \rangle = \sqrt{3} \lambda_{\mathrm{Mo}}/2, ~
\langle d_{xy} {\downarrow} | H_{\mathrm{SOI}} |  d_{xz} {\uparrow} \rangle = -i \lambda_{\mathrm{Mo}}/2, \\
& \langle d_{x^2 - y^2} {\downarrow} | H_{\mathrm{SOI}} |  d_{xz} {\uparrow} \rangle = - \lambda_{\mathrm{Mo}}/2, ~
\langle d_{z^2} {\downarrow} | H_{\mathrm{SOI}} |  d_{yz} {\uparrow} \rangle = i \sqrt{3} \lambda_{\mathrm{Mo}}/2, ~
\langle d_{xy} {\downarrow} | H_{\mathrm{SOI}} |  d_{yz} {\uparrow} \rangle = - \lambda_{\mathrm{Mo}}/2, \\
& \langle d_{x^2 - y^2} {\downarrow} | H_{\mathrm{SOI}} |  d_{yz} {\uparrow} \rangle = i \lambda_{\mathrm{Mo}}/2, ~
\langle d_{xz} {\downarrow} | H_{\mathrm{SOI}} |  d_{z^2} {\uparrow} \rangle = - \sqrt{3} \lambda_{\mathrm{Mo}}/2, ~
\langle d_{yz} {\downarrow} | H_{\mathrm{SOI}} |  d_{z^2} {\uparrow} \rangle = -i \sqrt{3} \lambda_{\mathrm{Mo}}/2, \\
& \langle d_{x^2-y^2} {\uparrow} | H_{\mathrm{SOI}} |  d_{xy} {\uparrow} \rangle = - i \lambda_{\mathrm{Mo}}, ~
\langle d_{xz} {\downarrow} | H_{\mathrm{SOI}} |  d_{xy} {\uparrow} \rangle = i \lambda_{\mathrm{Mo}}/2, ~
\langle d_{yz} {\downarrow} | H_{\mathrm{SOI}} |  d_{xy} {\uparrow} \rangle = \lambda_{\mathrm{Mo}}/2, \\
& \langle d_{xz} {\downarrow} | H_{\mathrm{SOI}} |  d_{x^2-y^2} {\uparrow} \rangle = \lambda_{\mathrm{Mo}}/2, ~
\langle d_{yz} {\downarrow} | H_{\mathrm{SOI}} |  d_{x^2-y^2} {\uparrow} \rangle = -i \lambda_{\mathrm{Mo}}/2, ~
\langle d_{yz} {\downarrow} | H_{\mathrm{SOI}} |  d_{xz} {\downarrow} \rangle =-i \lambda_{\mathrm{Mo}}/2, \\
& \langle d_{x^2 - y^2} {\downarrow} | H_{\mathrm{SOI}} |  d_{xy} {\downarrow} \rangle =i \lambda_{\mathrm{Mo}}.
\end{split}\]
The strength of the atomic SOC is $\lambda_{\mathrm{X}_A} = \lambda_{\mathrm{X}_B} = 0.0556$ eV for $\mathrm{X} = \mathrm{S}$, $0.2470$ eV for $\mathrm{X} = \mathrm{Se}$, and $\lambda_\mathrm{Mo} = 0.0836$ eV \cite{Fang2015}.

The physics of nodal topological superconductivity in monolayer \ce{MoX2} materials is governed by the low-energy dispersion of the spin-split lowest conduction band around the high symmetry points $\pm \bm K$.
The $11$-orbital tight-binding model captures the orbital character and symmetry of the monolayer bands and reproduces the main features of band structure over the entire Brillouin zone.
However, at the small energy scales close to the the conduction band minimum at $\bm K$, there are deviations in the $11$-orbital model from the first-principles calculations.
This is because the model is optimized to reflect the band structure over the entire Brillouin zone, but not to accurately capture nuances in the low-energy dispersion of individual bands near high-symmetry points, for which ${\bm k \cdot \bm p}$ Hamiltonians are generally more suitable.
Unlike the $11$-orbital model, the three-orbital model and the continuum model in the main part of this manuscript are both optimized to capture the relevant low-energy physics.
As a result, there are quantitative differences in the low-energy dispersion of the spin-split conduction band between the $11$-orbital model and the other two models, such as conduction band crossings that occur further away from the high-symmetry points $\bm K$.
This difference is intrinsic to the design of the $11$-orbital model~\cite{Fang2015}, and that despite the quantitative differences, all results obtained with the $11$-orbital model are in qualitative agreement with the continuum and three-orbital models.

\subsection{Tight-binding models for nodal topological superconductivity in \ce{MoX2}}
We model a superconducting monolayer \ce{MoX2} at the mean-field level with the Bogoliubov-de Gennes (BdG) Hamiltonian
\begin{equation}
H_{\mathrm{BdG}}(\boldsymbol{k}) = \begin{bmatrix}
H(\boldsymbol{k}) - \mu I_{2N} & -i \Delta \sigma_y \otimes I_{N} \\
i \Delta \sigma_y \otimes I_{N} & -H^*(\boldsymbol{-k}) + \mu I_{2N}
\end{bmatrix},
\label{eq:bdg}
\end{equation}
with $\mu$ the chemical potential, $\Delta$ the $s$-wave pairing, and where the normal-state Hamiltonian $H(\boldsymbol{k})$ refers to either the three-orbital model \eqref{eq:ham_3} or the $11$-orbital model  \eqref{eq:H_11}, for which $N=3$ and $N=11$, respectively.
In this work, we only consider values of $\mu$ that lie in the conduction band.

The BdG Hamiltonian Eq.~(\ref{eq:bdg}) has the intrinsic particle-hole symmetry
\begin{equation}
\mathcal{P} = \tau_x \otimes I_{2N} \mathcal K, \quad
\mathcal{P} H_{\mathrm{BdG}}(\bm k) \mathcal{P}^{-1} = -H_{\mathrm{BdG}}(-\bm k),
\label{eq:particle-hole}
\end{equation}
such that $\mathcal{P}^2 = 1$, where $\mathcal K$ is the complex conjugation
operator and $\tau_{x,y,z}$ are Pauli matrices that act in particle-hole space.
In the absence of magnetic fields, the BdG Hamiltonian furthermore has time-reversal symmetry $\mathcal{T} H_{\mathrm{BdG}}(\boldsymbol{k}) \mathcal{T}^{-1} = H_{\mathrm{BdG}}(-\boldsymbol{k})$, with $\mathcal{T} = i \tau_0 \sigma_y \otimes I_{N} \mathcal K$. Although a nonzero in-plane magnetic field breaks both $\mathcal{T}$ and the mirror symmetry $M_{xy}$ individually, the BdG Hamiltonian nevertheless remains symmetric to their product, and we can therefore define the effective time-reversal operator $\tilde{\mathcal{T}}$ that leaves the BdG Hamiltonian invariant, namely
\begin{equation}
\tilde{\mathcal{T}} = M_{xy} \mathcal{T}, \quad
\tilde{\mathcal{T}} H_{\mathrm{BdG}}(\boldsymbol{k}) \tilde{\mathcal{T}}^{-1} = H_{\mathrm{BdG}}(-\boldsymbol{k}),
\label{eq:GTRS}
\end{equation}
such that $\tilde{\mathcal{T}}^2 = 1$.
Combining $\mathcal{P}$ and $\tilde{\mathcal{T}}$, we therefore find that the BdG Hamiltonian Eq.~(\ref{eq:bdg}) has the chiral symmetry
\begin{equation}
\mathcal{C} = \tilde{\mathcal{T}} \mathcal{P}, \quad
\mathcal{C} H_{\mathrm{BdG}}(\boldsymbol{k}) \mathcal{C}^{-1} = -H_{\mathrm{BdG}}(\boldsymbol{k}), \quad
\mathcal{C}^2 = 1.
\label{eq:chiral}
\end{equation}

Since $\tilde{\mathcal{T}}^2=\mathcal{P}^2=1$, the BdG Hamiltonian \eqref{eq:bdg} describes a superconductor in class BDI, which in two-dimensions is a topologically trivial class.
Nevertheless, topologically protected flat bands of Andreev bound states may exist at the edges of such systems~\cite{Sato2011}.
To demonstrate this, we separate $\bm k = ({\bm k}_\parallel, {\bm k}_\perp)$ into two orthogonal projections, parallel ${\bm k}_\parallel$ and perpendicular ${\bm k}_\perp$ to a monolayer edge, respectively.
For example, an armchair edge of \ce{MoX2} is parallel to the $y$ direction and perpendicular to $x$, such that ${\bm k}_\parallel = k_y \hat{\bm y}$ and ${\bm k}_\perp = k_x \hat{\bm x}$.
Instead of applying the symmetry classification to the full two-dimensional BdG Hamiltonian $H_{\mathrm{BdG}}({\bm k})$, we reduce the dimension to one by treating ${\bm k}_\parallel$ as a parameter, and consider each one-dimensional Hamiltonian $H_{\mathrm{BdG}}({\bm k}) = H_{\mathrm{BdG}}^{{\bm k}_\parallel}({\bm k}_\perp)$ at a fixed ${\bm k}_\parallel$ separately.
Now, $\mathcal{P}$ and $\tilde{\mathcal{T}}$ are in general not symmetries of the one-dimensional Hamiltonian $H_{\mathrm{BdG}}^{{\bm k}_\parallel}({\bm k}_\perp)$ at a fixed ${\bm k}_\parallel$, because they flip the sign of both ${\bm k}_\parallel$ and ${\bm k}_\perp$.
Indeed, we generally find that no one-dimensional particle-hole $\mathcal{P}_{1\mathrm{D}}$ or time-reversal $\mathcal{T}_{1\mathrm{D}}$ type symmetries that satisfy $\mathcal{P}_{1\mathrm{D}} H_{\mathrm{BdG}}^{{\bm k}_\parallel}({\bm k}_\perp) \mathcal{P}_{1\mathrm{D}}^{-1} =-H_{\mathrm{BdG}}^{{\bm k}_\parallel}(-{\bm k}_\perp)$ or $\mathcal{T}_{1\mathrm{D}} H_{\mathrm{BdG}}^{{\bm k}_\parallel}({\bm k}_\perp) \mathcal{T}_{1\mathrm{D}}^{-1} =H_{\mathrm{BdG}}^{{\bm k}_\parallel}(-{\bm k}_\perp)$ exist in the tight-binding BdG models, even after performing a systematic search for such symmetries \cite{Varjas2018}.
The reason is that for the one-dimensional symmetries $\mathcal{P}_{1\mathrm{D}}$ and $\mathcal{T}_{1\mathrm{D}}$ to exist, there should be an extra unitary symmetry $V_\parallel$ commuting with the Hamiltonian that maps ${\bm k}_\parallel \rightarrow -{\bm k}_\parallel$.
We could then construct a $1\mathrm{D}$ symmetry $\mathcal{P}_{1\mathrm{D}}$ or $\mathcal{T}_{1\mathrm{D}}$ with the product of $V_\parallel$ and the corresponding $2\mathrm{D}$ symmetry.
However, for generic $({\bm k}_\perp, {\bm k}_\parallel)$, we find that no such symmetry $V_\parallel$ exists, and hence $\mathcal{P}_{1\mathrm{D}}$ and $\mathcal{T}_{1\mathrm{D}}$ are generally non-existent.
Regardless, the chiral symmetry \eqref{eq:chiral} is valid for any choice of directions in the Brillouin zone since $\mathcal{C}$ always leaves the momentum unchanged, \emph{i.e.} $\mathcal{C} H_{\mathrm{BdG}}^{k_\parallel}(\boldsymbol{k}_\perp) \mathcal{C}^{-1} = -H_{\mathrm{BdG}}^{ k_\parallel}(\boldsymbol{k}_\perp)$.
We therefore conclude that the one-dimensional Hamiltonians $H_{\mathrm{BdG}}^{{\bm k}_\parallel}({\bm k}_\perp)$ at a fixed ${\bm k}_\parallel$ belong to symmetry class AIII.

The topological number relevant for one-dimensional systems in class AIII is the winding number \cite{Schnyder2008, Sato2011}, which at the parallel momentum ${\bm k}_\parallel$ is given by
\begin{equation}
W(\boldsymbol{k}_\parallel) = \frac{1}{2\pi i} \int\limits_{\mathrm{BZ}}\frac{dz(\boldsymbol{k}_\perp)}{z(\boldsymbol{k}_\perp)}.
\label{eq:compute_winding}
\end{equation}
The integration is performed for a fixed value of $\boldsymbol{k}_\parallel$ over the one-dimensional Brillouin zone along the direction of $\boldsymbol{k}_\perp$, which is a closed loop.
Here, $z(\boldsymbol{k}_\perp) = \mathrm{det}\left\{ A(\boldsymbol{k}_\perp) \right\}/\left|\mathrm{det}\left\{ A(\boldsymbol{k}_\perp) \right\} \right|$ with
\begin{equation}
U_\mathcal{C}^\dagger H_{\mathrm{BdG}}(\boldsymbol{k}) U_\mathcal{C} = 
\begin{bmatrix}
0 & A({\boldsymbol{k}}) \\
A^{\dagger}({\boldsymbol{k}}) & 0
\end{bmatrix},
\end{equation}
and $U_\mathcal{C}$ the unitary matrix that diagonalizes $\mathcal{C}$, $U_\mathcal{C}^\dagger \mathcal{C} U_\mathcal{C} = \tau_z \otimes I_{2N}$.
The winding number is quantized to $W(\boldsymbol{k}_\parallel) \in \mathbb{Z}$, and changes only when the integration path over ${\bm k}_\perp$ intersects a nodal point \cite{Beri2010, Sato2011}, where the system is gapless such that $\mathrm{det}[A(\bm k)]=0$.
When $W(\boldsymbol{k}_\parallel)$ is nonzero, zero-energy states exist at the edge of the monolayer at the parallel momentum $\boldsymbol{k}_\parallel$.
Since $W({\bm k}_\parallel)$ only changes at values of ${\bm k}_\parallel$ where the integration path over ${\bm k}_\perp$ crosses a nodal point, $W(\boldsymbol{k}_\parallel)$ is generally nonzero in finite intervals of ${\bm k}_\parallel$, forming flat bands of Andreev bound states in the dispersion that are localized at the monolayer edge.

\subsection{Phase diagrams for $\text{MoX}_2$ monolayers and comparison with tight-binding calculations}
\begin{figure}[!tbh]
\renewcommand{\thesubfigure}{i}
\subfloat[\ce{MoTe2} ${\bm k} \cdot {\bm p}$ model.]{%
  \includegraphics[width=0.47\columnwidth]{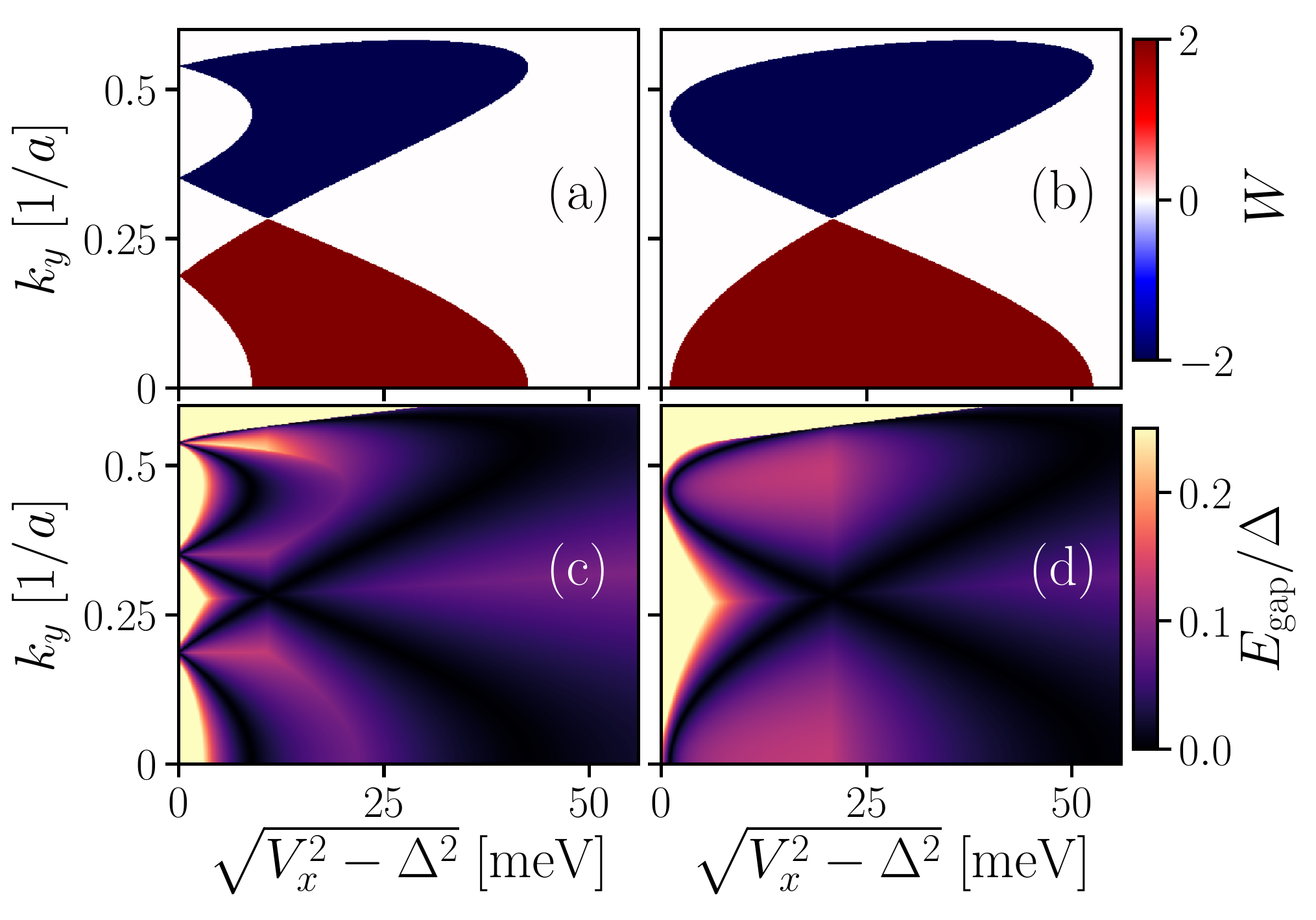}%
}\hfill
\renewcommand{\thesubfigure}{ii}
\subfloat[\ce{MoS2} ${\bm k} \cdot {\bm p}$ model.]{%
  \includegraphics[width=0.47\columnwidth]{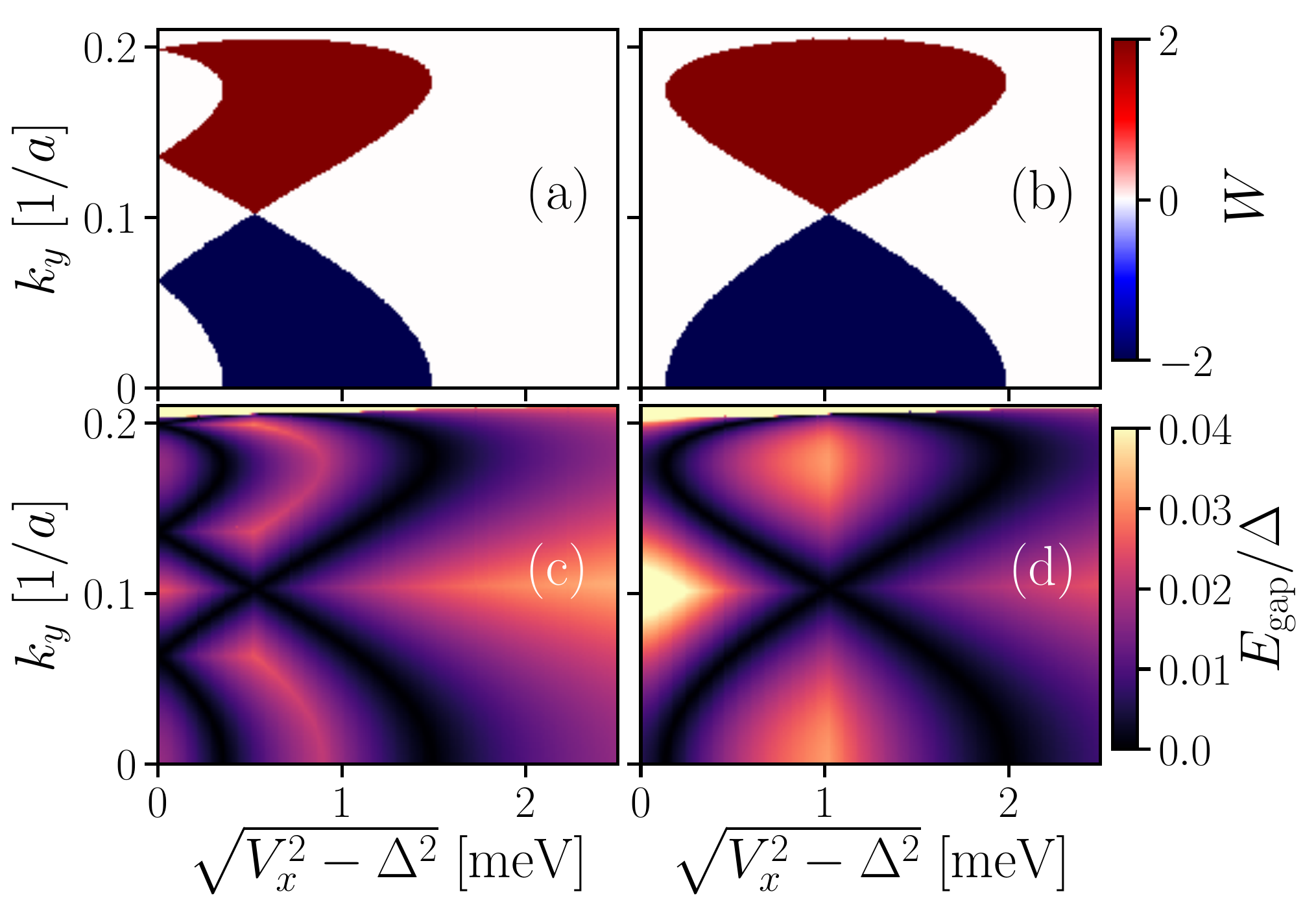}%
}
  \\
  \renewcommand{\thesubfigure}{iii}
\subfloat[\ce{MoS2} 3-orbital tight-binding model.]{%
  \includegraphics[width=0.47\columnwidth]{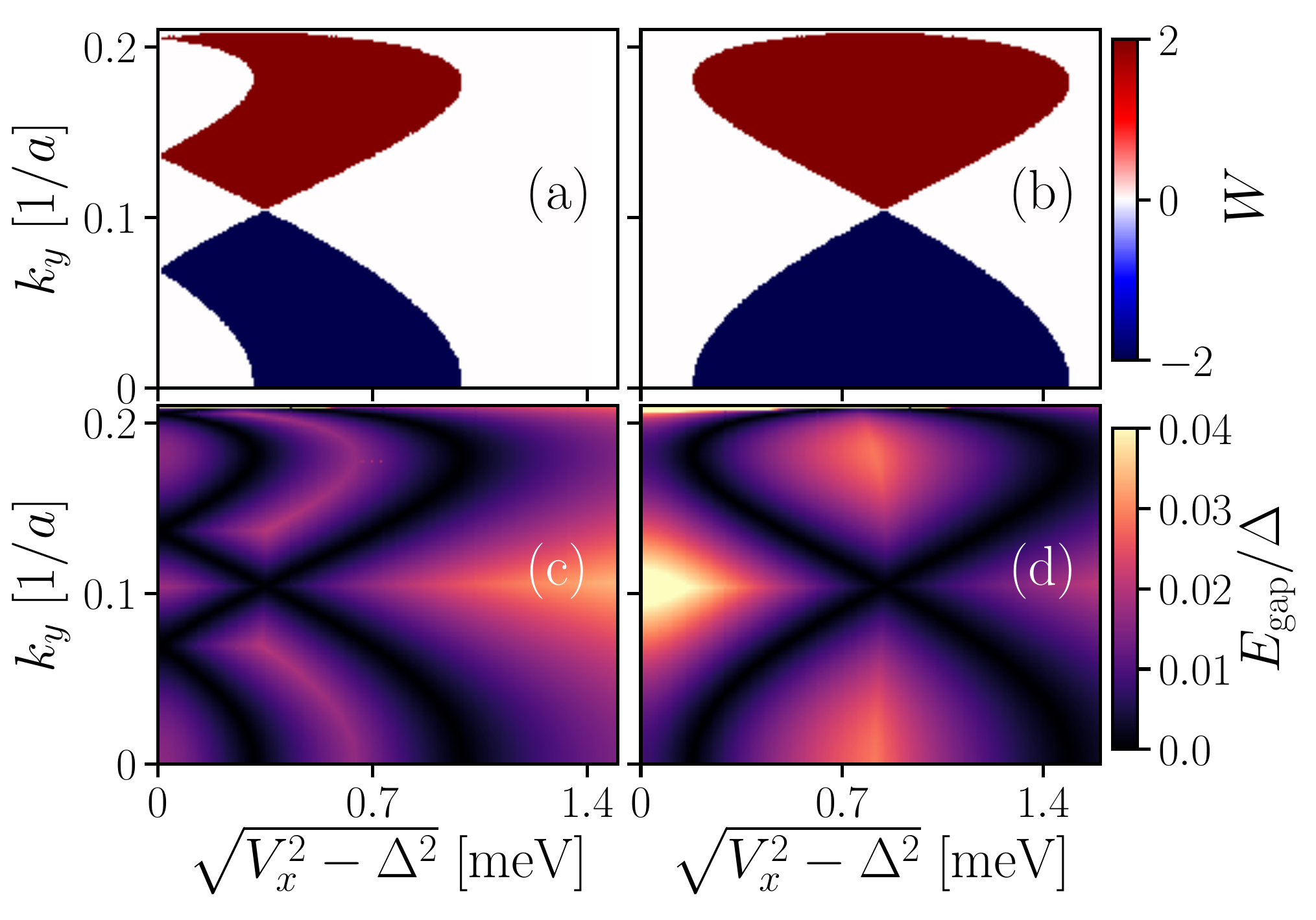}%
}\hfill
  \renewcommand{\thesubfigure}{iv}
\subfloat[\ce{MoS2} 11-orbital tight-binding model.]{%
  \includegraphics[width=0.47\columnwidth]{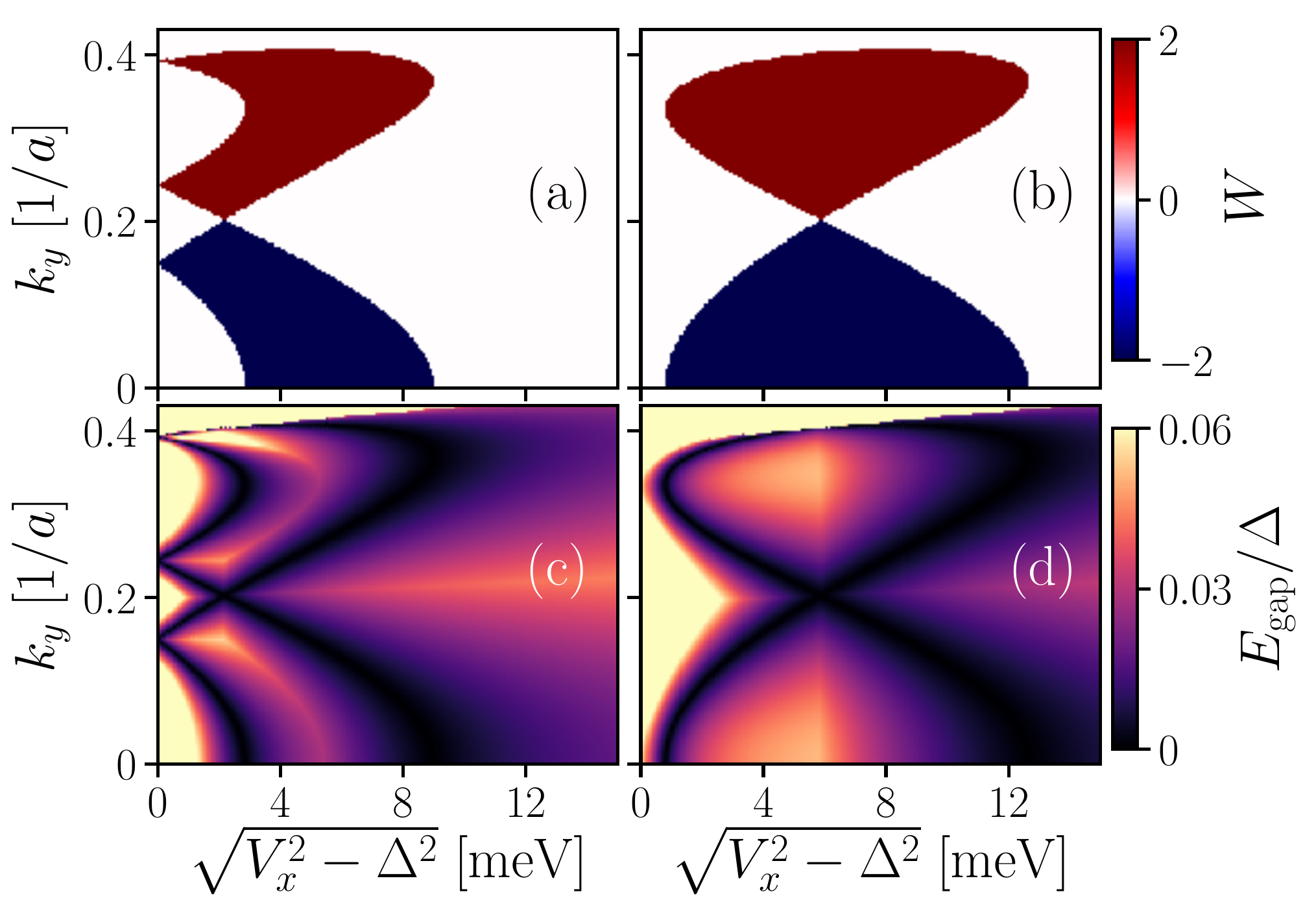}%
}
\caption{Topological phase diagrams and excitation gaps along the armchair direction ${\bm k}_\parallel = k_y \hat{\bm y}$ for (i) \ce{MoTe2} from the continuum ${\bm k} \cdot {\bm p}$ model, and (ii-iv) MoS2, comparing the ${\bm k} \cdot {\bm p}$ and three- and 11-orbital tight-binding models.
In all cases, panels (a) and (b) show the winding number $W$ computed using Eq.~\eqref{eq:compute_winding} as a function of $k_y$ and $\sqrt{V_x^2-\Delta^2}$ for (a) $\mu<\mu_1$ and (b) $\mu_1<\mu<(\mu_1+\mu_2)/2$, representative of regimes I and II of Fig.~\ref{fig:hex_nodal_points}(b) of the main text.
For $(\mu_1+\mu_2)/2<\mu<\mu_2$ and $\mu>\mu_2$, the phase diagrams are similar to (a) and (b) respectively, but with opposite winding numbers.
In all cases, panels (c) and (d) show the topological excitation gap $E_{\rm gap}$ corresponding to (a) and (b) separately.}
\label{fig:compare_topo}
\end{figure}
In Fig.~\ref{fig:compare_topo} (i) and (ii), we show examples of (a, b) topological phase diagrams and (c, d) maps of the corresponding topological excitation gap for superconducting monolayers of \ce{MoTe2} and \ce{MoS2}, respectively, obtained using the continuum ${\bm k} \cdot {\bm p}$ Hamiltonian.
Analogous data for monolayer \ce{MoSe2} is shown in Fig.~\ref{fig:winding_gap} of the main text.
We see that the phase diagrams of all three \ce{MoX2} materials are qualitatively similar, but with some quantitative differences, for instance a much larger excitation gap in \ce{MoTe2} than in \ce{MoS2}.
This is because the continuum Hamiltonians for all three materials are identical in form and have the same symmetries, see Eq.~\eqref{normal_H} of the main text.
Crucially, despite the differences in material parameters between the three types of monolayer, they all exhibit crossings between the spin-split conduction band where the SOC vanishes, as illustrated in Fig.~\ref{fig:crossings_11}, which makes the realization of the nodal topological phase possible.
The sign of the winding number for \ce{MoTe2} in (i) is opposite to that of \ce{MoS2} in (ii) because spin-polarization of the conduction bands is inverted [see Table \ref{tab:kp_material}].
Figures \ref{fig:compare_topo} (iii) and (iv) show analogous results for \ce{MoS2} obtained using (iii) the three-orbital tight-binding model, and (iv) the 11-orbital tight-binding model.
The three-orbital tight-binding model agrees well with the continuum model, but there are more prominent quantitative differences with the 11-orbital model, because the latter is not optimized to accurately describe the low-energy physics near the $\bm K$ valleys \cite{Fang2015}.
Nonetheless, all three models are in qualitative agreement.

\begin{figure}[!tbh]
\renewcommand{\thesubfigure}{i}
\subfloat[11-orbital tight-binding model.]{%
  \includegraphics[width=0.47\columnwidth]{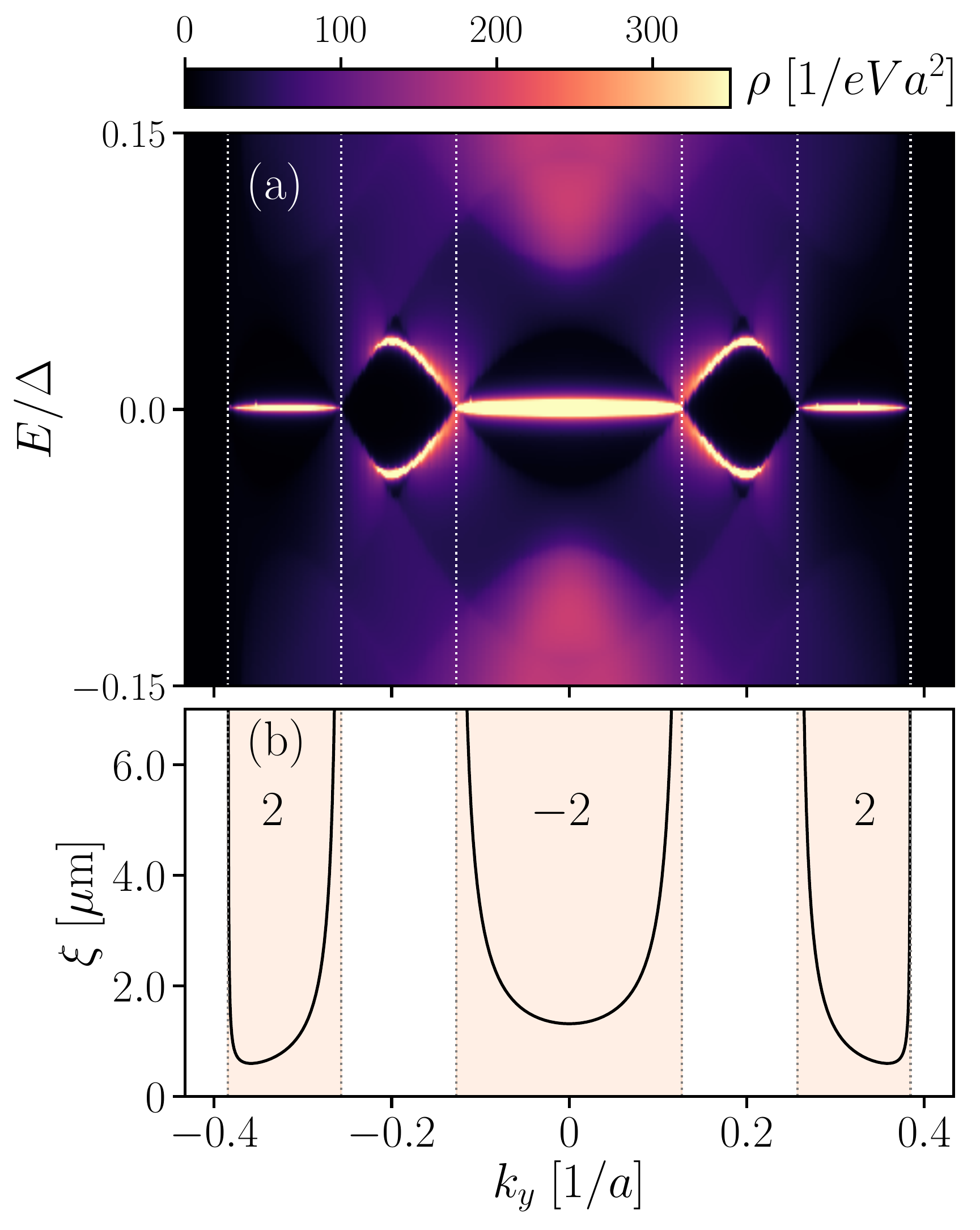}%
}\hfill
\renewcommand{\thesubfigure}{ii}
\subfloat[3-orbital tight-binding model.]{%
  \includegraphics[width=0.47\columnwidth]{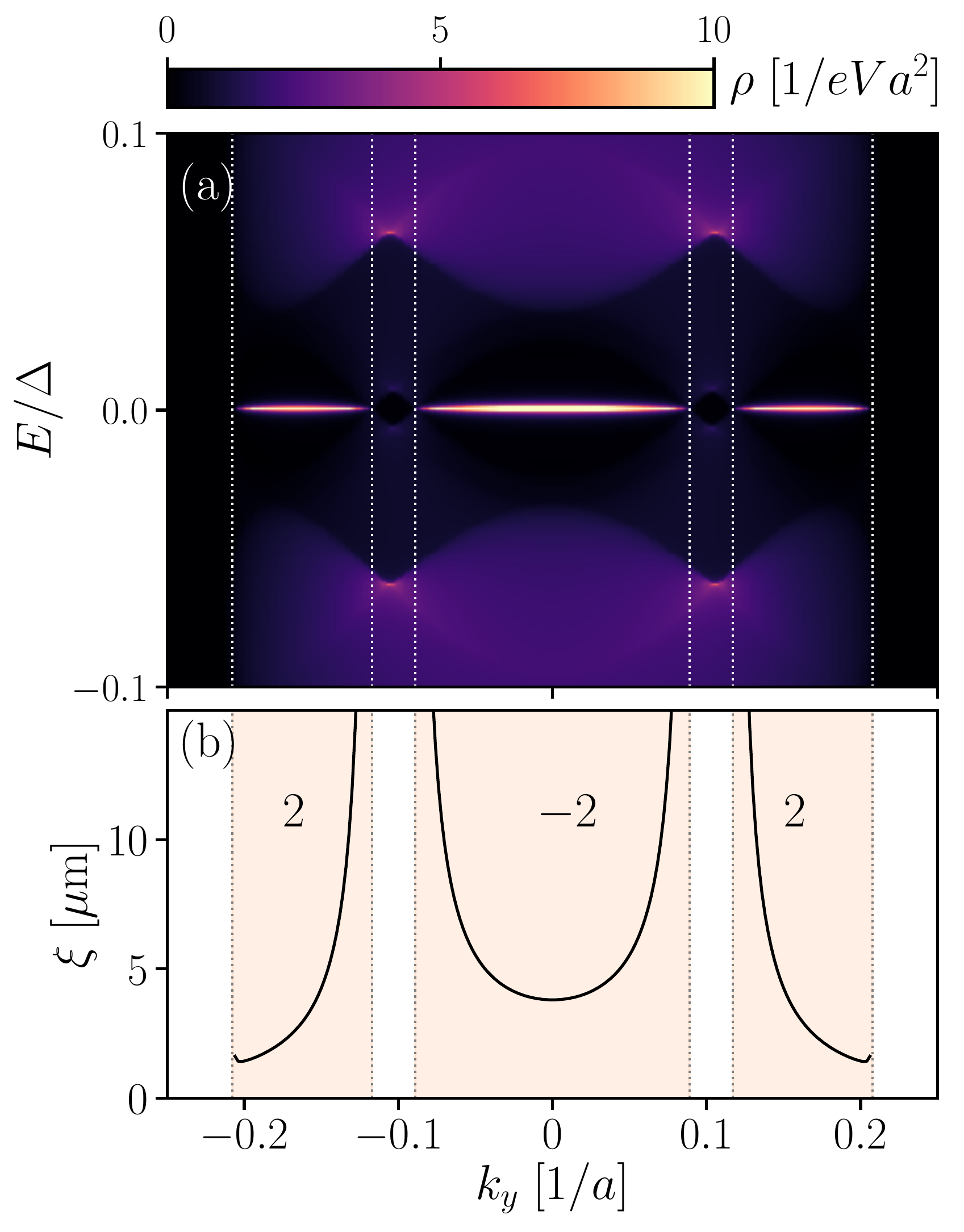}%
}
\caption{(a) Local density of states and (b) decay length of the nontrivial edge states at an armchair edge of a monolayer \ce{MoS2} obtained using the (i) 11-orbital tight-binding model, and (ii) the three-orbital tight-binding model.
In both cases $\Delta = 0.8$ meV, with other parameters in regime II of of Fig.~1(b) of the main text.
When the winding number is nonzero, flat bands of zero-energy Andreev edge states connecting the edge projections of the nodal points appear at the armchair edge.
The nontrivial phases are marked by the shaded regions in (b) with the nonzero winding numbers in the insets.
We have $\mu = 1.830$ eV and $V_x = 3.105$ meV in (i), and $\mu=\SI{1.709}{eV}$, $V_x=\SI{1.131}{meV}$ in (ii).}
\label{fig:compare_DOS}
\end{figure}
Figure \ref{fig:compare_DOS} shows the local density of states at an armchair edge of a superconducting monolayer \ce{MoS2} using parameters within regime II of Fig.~\ref{fig:hex_nodal_points}(b) of the main text, obtained from the (i) 11-orbital tight-binding model, and the (ii) three-orbital tight-binding model.
We see flat bands of zero-energy Andreev bound states that are localized at the edge manifest in regions where the winding number $W(k_y)$ is nonzero.
In the parameter regime II we consider here, there are $6$ nodal points near each inequivalent $K$ valley.
The projections of the nodal points onto the armchair edge partition the one-dimensional Brillouin zone of the armchair edge into $7$ segments, with the flat bands appearing in segments where the winding number is nonzero.
Despite quantitative differences between the two models, the figures are in qualitative agreement.

We conclude this section with a brief comparison of the two tight-binding models.
Both models agree with our analysis of the nodal topological phase based on the continuum Hamiltonian in the main text.
Since it contains fewer basis orbitals per unit cell, the three-orbital model is less cumbersome to work with than the $11$-orbital model.
This also makes the three-orbital model more suitable for performing large-scale simulations of finite systems with multiple unit cells, as the size of the Hamiltonian will scale better than using the $11$-orbital model.
In addition, the three-orbital model is more accurate than the $11$-orbital model near the high symmetry points $\pm \bm K$ which are most relevant in our study, since the $11$-orbital model is designed to approximate the band structure in the entire Brillouin zone instead of only near $\pm \bm K$.
However, the three-orbital model only includes orbitals on the M atoms but not the X atoms, while the $11$-orbital model includes orbitals on all the atoms in the monolayer unit cell.
This makes the $11$-orbital model better suited than the three-orbital model to simulate effects that depend on the three dimensional structure of the monolayer, such as the orbital effects of a magnetic field.
Similarly, atomic SOC terms are sufficient to reproduce the crossings in the conduction band necessary for the nodal topological phase within the $11$-orbital model.
In the three-orbital model however, it is necessary to supplement the SOC on the M atom with an extra term due to virtual hoppings to orbitals on the X atoms, as shown in Section \ref{section:3orb}.

\subsection{Topological phases for arbitrary edge cuts}
The presence of zero-energy flat bands for ribbons with different edge orientation can be predicted due to bulk-edge correspondence from the knowledge of the bulk topological invariant.
The usual way to probe the topological phase diagram is to write effective Hamiltonians for specific edge orientation.
Zero-energy states exist whenever the effective edge Hamiltonian has a non-trivial winding number.
This procedure is cumbersome since it requires redefining for each edge a different Hamiltonian, which depending of the edge orientation, might be represented by an unwieldy large matrix.

An alternative method developed in Ref.~\cite{Delplace2011}, allows us to keep the bulk Hamiltonian unchanged, but instead vary an effective Brillouin zone.
As described in the main text, we assume that the lattice termination edge forms a one-dimensional superlattice, with a translation period given by the superlattice vector
\begin{equation}
\bm T=m\bm a_1+n \bm a_2,
\end{equation}
which is parallel to the lattice termination edge.
Here, $m$ and $n$ are coprime integers, and the Bravais lattice vectors are given in \eqref{eq:lat_vecs}.

Along the edge, parallel to $\bm T$, we define the conserved momentum $k_\parallel$, with values in the 1D Brillouin zone of size $\Delta k_\parallel=2\pi/|\bm T|$, namely
\begin{equation}
k_\parallel \in \Delta k_\parallel
=\frac{\pi}{a}[-\frac{1}{\sqrt{m^2+n^2+mn}},
\frac{1}{\sqrt{m^2+n^2+mn}}). \label{eq:parBZ}
\end{equation}
The momentum span $\Delta k_\perp$ in the direction perpendicular to the edge is constrained such that the area of Brillouin zone is conserved,
\begin{equation}
k_\perp \in \Delta k_\perp = \frac{2\pi}{\sqrt 3 a}
[-\sqrt{m^2+n^2+mn},\sqrt{m^2+n^2+mn}), \label{eq:perpBZ}
\end{equation}
with the momentum $k_\perp$ perpendicular to the edge.
For a given lattice termination boundary characterized by $\bm T$, the winding number at the momentum $k_\parallel$ in the 1D Brillouin zone \eqref{eq:parBZ} follows from Eq.~\eqref{eq:compute_winding} by integrating over all $k_\perp$ in \eqref{eq:perpBZ}.
Note that \eqref{eq:perpBZ} is exactly one period in reciprocal space, and the integral over $k_\perp$ to compute the winding number is therefore over a closed loop.

\subsection{Orbital effect of the in-plane magnetic field}
In this section, we discuss the orbital effects of the in-plane magnetic field on the nodal topological superconducting phase in the monolayer \ce{MoX2}.
For simplicity, we assume a magnetic field ${\bm B} = B \hat{\bm x}$ along $x$ only.
To preserve translational invariance in the monolayer plane, we pick the vector potential ${\bm A} = -Bz\hat{\bm y}$.

Including the orbital effect of the magnetic field does not alter the symmetry classification of the BdG Hamiltonian \eqref{eq:bdg}.
To demonstrate this, we include the magnetic field in the normal state continuum model with the kinetic momentum substitution $k_y \rightarrow \hbar k_y + eA = k_y - eBz$, with $e$ the unit charge.
Since $M_{xy}: z \rightarrow -z$ and $\mathcal{T}: {\bm k} \rightarrow -{\bm k}$, we see that the new kinetic momentum transforms identically under the product $M_{xy} \mathcal{T}$ with and without the orbital effects of the magnetic field.
Thus, the normal-state Hamiltonian with orbital effects included remains invariant to the generalized time-reversal symmetry $\tilde{\mathcal{T}} = M_{xy} \mathcal{T}$, and the BdG Hamiltonian therefore also to the chiral symmetry $\mathcal{C}$.
We have verified this numerically by including the orbital effects of the magnetic field in the 11-orbital tight-binding model, with a Peierls substitution for the hopping matrices $T_{\alpha \beta}^{(j)} \rightarrow T_{\alpha \beta}^{(j)} \exp{(-i\frac{e}{\hbar} \int \boldsymbol{A} \cdot d\boldsymbol{r} )}$ \cite{Peierls1933, Hofstadter1976, Boykin2001}.
We find negligible quantitative differences in our numerical results with and without the orbital effects included and no qualitative differences, and therefore neglect the orbital effects in our calculations.
This is reasonable, because the magnetic length $l_B = \sqrt{\hbar/eB}$ with $\hbar$ the reduced Planck's constant is in the nanometers even up to extremely large fields $\lesssim 100$ T, and thus always much larger than the separation $d \approx 3~\mathrm{\AA}$ between the top and bottom layers of X atoms in the monolayer.

\end{document}